\documentclass[twocolumn,aps,prl,superscriptaddress,amsmath,amssymb]{revtex4-2}
\usepackage{amssymb}
\usepackage{amsmath}
\usepackage{amsfonts}
\usepackage{graphicx}
\usepackage{color}
\usepackage{bm}
\usepackage[colorlinks,linkcolor=blue,anchorcolor=blue,citecolor=blue,urlcolor=blue]{hyperref}

\begin{document}

\title{The emergence of net chirality in two-dimensional Dirac fermions system with altermagnetic mass}

\author{Peng-Yi Liu}
\thanks{These authors contributed equally to this work.}
\affiliation{International Center for Quantum Materials, School of Physics, Peking University, Beijing 100871, China}

\author{Yu-Hao Wan}
\thanks{These authors contributed equally to this work.}
\affiliation{International Center for Quantum Materials, School of Physics, Peking University, Beijing 100871, China}

\author{Qing-Feng Sun}
\email[Corresponding author: ]{sunqf@pku.edu.cn}
\affiliation{International Center for Quantum Materials, School of Physics, Peking University, Beijing 100871, China}
\affiliation{Hefei National Laboratory, Hefei 230088, China}

\begin{abstract}
In two-dimensional lattice systems, massless Dirac fermions undergo doubling, leading to the cancellation of net chirality.
We demonstrate that the recently discovered altermagnetism can induce a unique mass term, the altermagnetic mass term, which gaps out Dirac cones with one chirality while maintaining the other gapless, leading to the emergence of net chirality.
The surviving gapless Dirac cones retain identical winding numbers and exhibit the quantum anomalous Hall effect in the presence of the trivial constant mass term.
When subjected to an external magnetic field, the altermagnetic mass induces Landau level asymmetry in Dirac fermions, resulting in fully valley-polarized quantum Hall edge states.
Our findings reveal that Dirac fermions with the altermagnetic mass harbor rich physical phenomena warranting further exploration.
\end{abstract}

\maketitle

In condensed matter physics, emergent quasiparticle excitations with linear dispersions in materials provide a unique platform for exploring relativistic quantum phenomena \cite{graphene_2006_Klein,graphene_2006_zit,graphene_2013_ACS}.
Two-dimensional (2D) systems, exemplified by graphene and the surface states of three-dimensional topological insulators, successfully realize massless Dirac fermions (DFs) whose low-energy excitations exhibit gapless linear dispersion near the Dirac points \cite{graphene_2005_1,graphene_2005_2,graphene_2009,topological_2010,topological_2008a,topological_2008b}.
When DFs acquire mass, such as the constant Dirac mass, the chiral symmetry is broken, accompanied by the opening of a gap \cite{graphene_2009,mass_2024}.
DFs with mass leading to rich physics, such as the quantum anomalous Hall (QAH) and quantum spin Hall effect \cite{parity_1988,mass_2005,topological_2008b}.
Furthermore, massless DFs with a single flavor cannot exist independently in 2D lattice systems due to the Nielsen-Ninomiya theorem \cite{NN_1981a,NN_1981b,NN_1983}.
Interestingly, introducing a mass term quadratic in momentum, known as the Wilson mass \cite{wilson_1982,wilson_2022}, circumvents Fermi doubling and induces a parity anomaly \cite{parity_1983,parity_1984,parity_1988,parity_2022a,parity_2024a,parity_2024b,parity_2025} with half-quantized Hall conductivity.
However, such systems remain experimentally unrealized in real materials \cite{wilson_2023}, because of the difficulty in achieving momentum-dependent mass.

Recently, altermagnetism has emerged as a novel magnetic state distinct from ferromagnetism and antiferromagnetism \cite{altermagnet_2020,altermagnet_2021,altermagnet_2022}.
This state breaks time-reversal symmetry without exhibiting macroscopic magnetism, offering significant application potential in spintronics \cite{altermagnet_2022_split,altermagnet_2022_split2,altermagnet_2022_GMR,addr1,addr2,addchang,new_alterweyl,new_alterTI}.
Intriguingly, the altermagnets split the spin by introducing a momentum-dependent Zeeman term, which can break the chiral symmetry and be used as a momentum-dependent mass.

In this Letter, we identify that the influence of altermagnetism on 2D DFs can be interpreted as a unique momentum-dependent mass term, named the altermagnetic mass term.
Remarkably, while the exact chiral symmetry is broken by the altermagnetic mass, the two gapless Dirac cones maintain the identical chiral-like winding number.
In the gapped regime, these Dirac cones exhibit quantized Chern numbers and Hall conductivity.
Furthermore, under an applied magnetic field, the altermagnetic mass induces a Landau level (LL) asymmetry, resulting in fully valley-polarized edge states.

\begin{figure}
	\centering
	\includegraphics[width=\columnwidth]{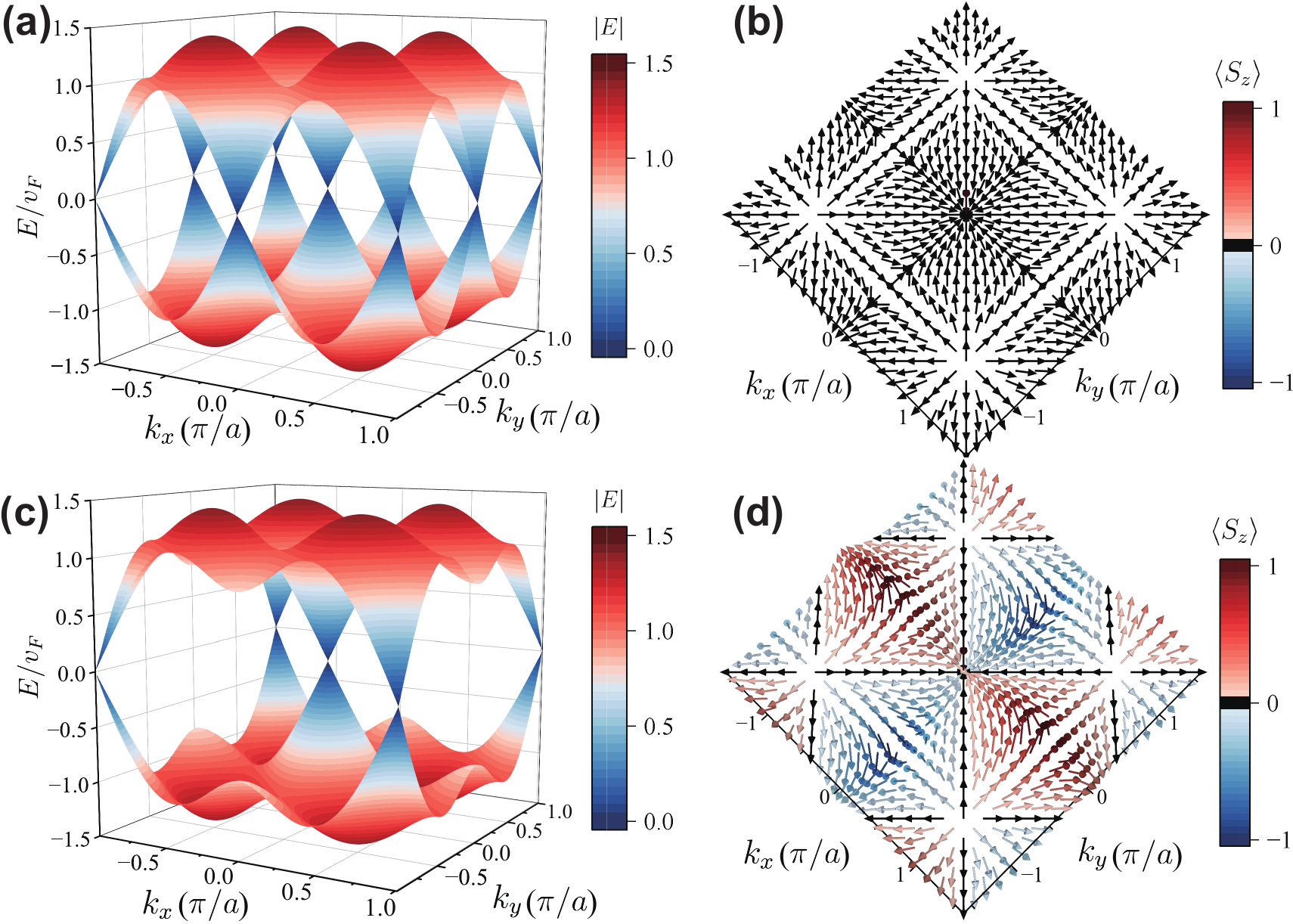}
	\caption{DFs without [(a) and (b)] and with [(c) and (d)] the altermagnetic mass in a 2D lattice. (a) and (c) show the band structures of $H_{\rm{DF}}$ and $H_{\rm{DFA}}$, respectively, with the color reflecting the absolute value of energy.
	(b) and (d) are the corresponding spin textures $\bm{S}$ of the negative-energy state, with the color reflecting the $S_z$ component.
	The unit of energy is ${v_F}$ and $m_{\rm Am}=0.4{v_F}$ is maintained throughout the paper.
	}
	\label{fig 1}
\end{figure}

For general two-component DFs in 2D continuous systems, the Hamiltonian leads to the linear dispersion and is given by $\sum_{i,j=x,y}k_i v_{ij} \sigma_j$, where $\sigma_j$ is the Pauli matrix in the $j$-direction of the spin space and $v$ is a $2\times 2$ matrix.
Because this Hamiltonian anticommutes with $\sigma_z$, the system has chiral symmetry \cite{topological_2010}, and when $v_{ij}=v_F\delta_{ij}$, the chirality is given by $\chi={{\rm{sgn}}[{\rm{det}}(v)]}=+1$ \cite{topological_2012}.
However, when introduced into lattice systems, net chiralities cannot exist, due to the Nielsen-Ninomiya theorem \cite{NN_1983}.
The Hamiltonian of the DF discretized by a square lattice is $H_{\rm{DF}}(\bm{k})=\frac{v_F}{a}[\sin (k_x{a}) \sigma_x +\sin (k_y{a}) \sigma_y]$ with the lattice constant $a=1$ ($a$ is retained in some places for clarity).
As a result of lattice regulation, the energy spectrum becomes ultraviolet complete, and four Dirac cones appear at $\Gamma(0,0)$, $X(\pi/a,0)$, $Y(0,\pi/a)$, and $M(\pi/a,\pi/a)$ points in the first Brillouin zone, known as the Fermi doubling, as shown in Fig. \ref{fig 1}(a).
The chirality of the four Dirac cones can be visualized by the spin texture, which is the expectation of the spin operator in the negative-energy state $\bm{S=\langle \sigma \rangle}$, as shown in Fig. \ref{fig 1}(b), and the chirality $\chi$ is equal to the winding number $W$ of the spin texture (see Sec. S1 of the Supplemental Material \cite{supp,winding_2007,winding_2021}).
At all positions in the Brillouin zone, $S_z=0$, because the Hamiltonian maintains chiral symmetry.
At $\Gamma$ and $M$, clockwise around the Dirac point, the spin texture also rotates clockwise, corresponding to $W=\chi=+1$ \cite{supp,winding_2007,winding_2021}.
In contrast, at $X$ and $Y$, the spin texture rotates anticlockwise, and $W=\chi=-1$.
The chiral pairwise cancellation of all four Dirac cones results in zero net chirality without quantum anomalies \cite{NN_1983}, just as in graphene.

According to the symmetry, \(d\)-wave altermagnetism can introduce \(m_{\rm Am}(k_x^2-k_y^2)\sigma_z\) terms into the Hamiltonian \cite{altermagnet_2022,altermagnet_2022_GMR,altermagnet_2025}.
We refer to this unique Zeeman term as the altermagnetic mass, which respects the symmetry of fourfold space rotation (\(\mathcal{C}_4\)) and time-reversal (\(\mathcal{T}\)).
Such $d$-wave altermagnetic configurations have been proposed in realistic materials, including $\rm RuO_2$, based on \emph{ab initio} calculations and ARPES experiments \cite{altermagnet_2020, RuO22,RuO23}.
For comparison, the trivial Dirac mass (\(m\sigma_z\)) opens a gap for all momenta, while the Wilson mass (\(m_{\text{Wilson}}k^2\sigma_z\)) opens a gap for all momenta except \(\Gamma\) point.
In contrast, the altermagnetic mass $m_{\rm Am}$ opens a gap for momenta except along the lines \(k_x=\pm k_y\).
In the discretized lattice, the Hamiltonian of the DF system with altermagnetic mass (DFA) changes to $H_{\rm{DFA}}(\bm{k})=H_{\rm{DF}}(\bm{k})+\frac{m_{\rm Am}}{a^2}[\cos (k_x{a}) -\cos (k_y{a})]\sigma_z$.
The discretization makes the appearance of a DFA at $\Gamma$ accompanied by a DFA with opposite altermagnetic mass at $M$, as shown in Fig. \ref{fig 1}(c).
Fig. \ref{fig 1}(d) shows the spin texture of two DFAs.
Although the altermagnetic mass breaks the chiral symmetry and introduces $S_z$ components, near the Dirac points, chiral-like characteristics are maintained, namely the winding number $W=+1$ for both $\Gamma$ and $M$ points.
A similar example is the Wilson fermion, which avoids fermion doubling by introducing mass $m_{\text{Wilson}}$ breaking the chiral symmetry \cite{wilson_1982}.
The difference is that Wilson fermions can be the transition between Chern insulators and normal insulators, and have a half-integer Hall conductance \cite{parity_2022a,wilson_2022}.
In contrast, the altermagnetic mass has $\mathcal{C}_4\mathcal{T}$ symmetry, with two Dirac cones contributing zero to the Chern number, resulting in the absence of linear Hall conductance.
Since the Dirac cones at $\Gamma$ and $M$ have the same winding in spin texture, once these cones are gapped by a trivial mass term \(m\sigma_z\), chiral features including the edge transport and QAH effect can emerge, as shown below.

To study the edges of DFAs, we calculate the band structure of a nanoribbon with a width in the $y$-direction $L_y=100a$ (open boundary condition) and infinite extent in the $x$-direction.
The Hamiltonian of nanoribbons is established by the Fourier transformation of $H_{\rm DFA}$ (see Sec. S2 of the Supplemental Material \cite{supp}).
As shown in Fig. \ref{fig 2}(a), the Dirac cone at $\Gamma$ ($M$) point is projected to the center (boundary) of the one-dimensional Brillouin zone.
More precisely, we zoom in on the band structure near the Dirac point at $k_x=0$ and the color represents the center of wavefunction in the $y$-direction $\langle y \rangle$ for each state. As shown in Fig. \ref{fig 2}(b),
the electron-type and hole-type bands nearly touch at the Dirac point, with linear dispersion.
The centers of wavefunction of states within the bulk bands are located at the center of the nanoribbon $\langle y \rangle\approx 0.5L_y$,
whereas the outermost states exhibit non-central spatial distributions, indicating them being edge states.
Specifically, edge states with positive (negative) velocities tend to be distributed towards smaller (larger) $y$-values.
However, these states are not exponentially decaying edge states; rather, as illustrated in Fig. \ref{fig 2}(c), which shows the squared modulus of the wavefunction $|\psi|^2$ in the $y$-direction.
They exhibit chiral metallic behavior with power-law decay from the edges \cite{powerlaw_2022}.

\begin{figure}[!htb]
	\centerline{\includegraphics[width=\columnwidth]{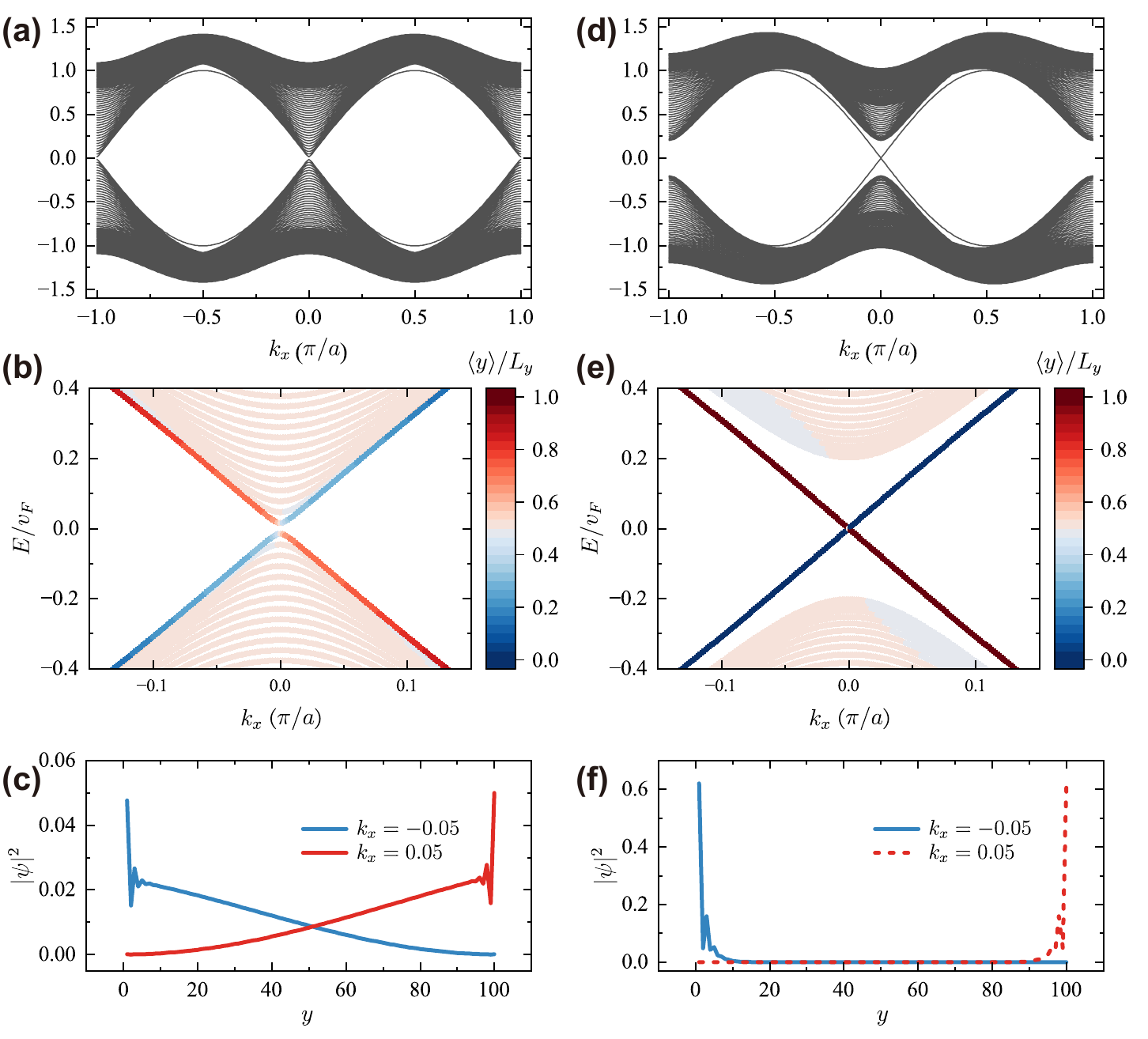}}
	\caption{The band structure of gapless (a-c) and gapped (d-f, $m=-0.2{v_F}$) DFAs nanoribbons in the $x$-direction, with the width in the $y$-direction $L_y=100a$.
	(a) and (d) are complete energy bands.
	(b) and (e) are enlarged views of the Dirac cone near $\Gamma$ point, with the color reflecting the center of wavefunctions in the $y$-direction $\langle y \rangle$.
	(c) and (f) are the modular squares of the wavefunctions of the outermost states at $k_x=\pm0.05$.
	}
	\label{fig 2}
\end{figure}

In the gapless state, the Chern number $\mathcal{C}$ exhibits a singularity, but when the system is a gapped state, $\mathcal{C}$ is well-defined and quantized.
Therefore, we expect that the property of two DFAs with the same winding number will be present in the gapped regime.
A Dirac cone with a trivial mass term (Zeeman term) $m\sigma_z$ contributes a half-integer Chern number, with the sign determined by the chirality and the sign of $m$, $\mathcal{C}={\rm sgn[det(}v)m]/2=\chi {\rm sgn(}m)/2$ \cite{topological_2012,topological_2013}.
According to this, conventional gapped fermion doubling systems, such as graphene, exhibit trivial $\mathcal{C}=0$ and no Hall conductance, as a result of chirality cancellation.
However, for gapped DFAs in lattice systems, although the chiral symmetry is no longer strictly maintained, the same winding number implies that it has a quantized Chern number $\mathcal{C}=W {\rm sgn(}m)$, which can be confirmed by examining its edge states.

Considering a nanoribbon of DFAs gapped by a Zeeman term, $H_{\rm{DFA}}+m\sigma_z$, Fig. \ref{fig 2}(d) illustrates that Dirac cones at both the center and edges of the Brillouin zone are gapped.
More importantly, within the energy gap near $k_x= 0$, chiral in-gap edge states emerge, as a result of bulk-edge correspondence \cite{topological_2012,topological_2013}.
These edge states emerge from the transition of the chiral metallic states depicted in Fig. \ref{fig 2}(b,c), manifesting the chiral characteristic of the DFAs.
This topological transition is governed by the altermagnetic mass, as evidenced by the absence of topological edge states in conventional fermion doubling systems, even when the gap is opened by $m\sigma_z$ (see Fig. S1 and Sec. S3 in the Supplemental Material \cite{supp}).
Fig. \ref{fig 2}(e) displays $\langle y\rangle$ of states near the Dirac point, showing that bulk states are nearly centered within the nanoribbon, while in-gap edge states exhibit significant localization at the edges of the nanoribbon.
We further present the wavefunctions of the edge states in Fig. \ref{fig 2}(f), which are localized at the edges of the nanoribbon with exponential decay, approaching zero at the center.

\begin{figure}[!htb]
	\centerline{\includegraphics[width=\columnwidth]{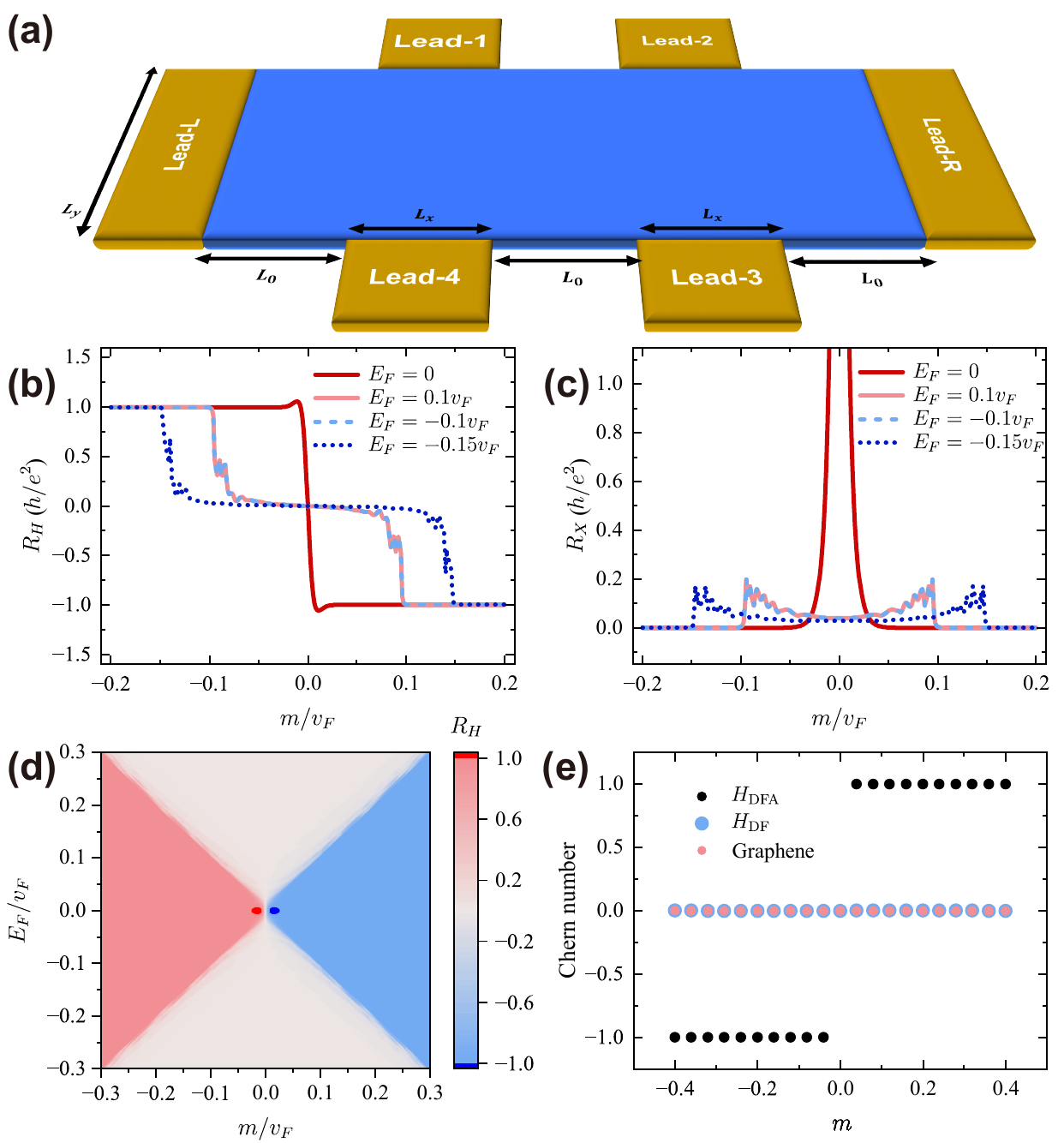}}
	\caption{The quantum anomalous Hall effect of DFAs.
		(a) Schematic of a Hall bar with six leads (shown in gold), where the blue region is composed of Dirac materials with altermagnetic mass and Zeeman term, described by $H_{\rm{DFA}}+m\sigma_z$.
		The width of lead-L and R is $L_y=100a$, the width of lead-(1-4) is $L_x=50a$, and the spacing between the leads is $L_0=50a$.
		(b,c) The Hall resistances (b) and longitudinal resistances (c) versus $m$ with four different $E_F$.
		(d) The phase diagram ($R_H$ in the unit of $h/e^2$) with $E_F$ and $m$.
		(e) The comparison of the Chern number under a Zeeman term $m\sigma_z$ of $H_{\rm DFA}$, $H_{\rm DF}$, and graphene.
		The energy unit of $m$ of DFA and DF is ${v_F}$, and that of graphene is the nearest-neighbor hopping.
	}
	\label{fig 3}
\end{figure}

The chiral edge states demonstrated above and the Chern numbers
suggest the presence of QAH in DFAs in the gapped regime.
To verify this, we model electron transport in a six-terminal Hall bar, illustrated in Fig. \ref{fig 3}(a), using the Landauer-B\"{u}ttiker formalism,
$I_p=\sum_q T_{pq} (V_p-V_q)$.
Here, $I_p$ and $V_p$ represent the current and voltage of lead-$p$, respectively, with $p,q=L,R,1,2,3,4$.
Through the non-equilibrium Green's function, the transmission coefficients $T_{pq}={\rm Tr}[\mathbf{\Gamma}_p \mathbf{G}^r_{pq} \mathbf{\Gamma}_q \mathbf{G}^a_{qp}]$ \cite{Green_1992,Green_2009,Peierls_2008,Peierls_2024,new_metalHall},
where $\mathbf{G}^r$ and $\mathbf{G}^a=(\mathbf{G}^r)^\dagger$ are the retarded and advanced Green's function, respectively.
With the Dyson equation, $\mathbf{G}^r=[(E_F+{\rm i}0^+)\mathbf{I}-\mathbf{H}_{\rm cen}-\sum_p\mathbf{\Sigma}_p^r]^{-1}$,
where $\mathbf{I}$ and $\mathbf{H}_{\rm cen}$ are the identity matrix and the Hamiltonian matrix of the center region [the blue region in Fig. \ref{fig 3}(a)], respectively.
$\mathbf{\Sigma}^r_p$ is the retarded self-energy caused by the coupling of lead-$p$.
The linewidth function of lead-$p$ is defined as $\mathbf{\Gamma}_p={\rm i}[\mathbf{\Sigma}^r_p-(\mathbf{\Sigma}^r_p)^\dagger]$.
For numerical simplicity, we adopt $\mathbf{\Sigma}^r_p=-\frac{\rm i}{2} v_F \mathbf{I}_p$, where $\mathbf{I}_p$ is the identity matrix of the sites attached to lead-$p$.
Leads-$L$ and -$R$ serve as the source and drain ($V_L=V_0$, $V_R=0$), while the remaining four leads serve as voltage probes ($I_{1,2,3,4}=0$).
Solving the above equations yields the probe voltages $V_{1,2,3,4}$ and current $I_L=-I_R$.
At last,
the Hall and longitudinal resistances are given by, $R_H=-1/\sigma_{xy}=(V_1-V_4)/I_L$ and $R_X=(V_1-V_2)/I_L$, respectively. 

\begin{figure*}[!htb]
	\centerline{\includegraphics[width=2\columnwidth]{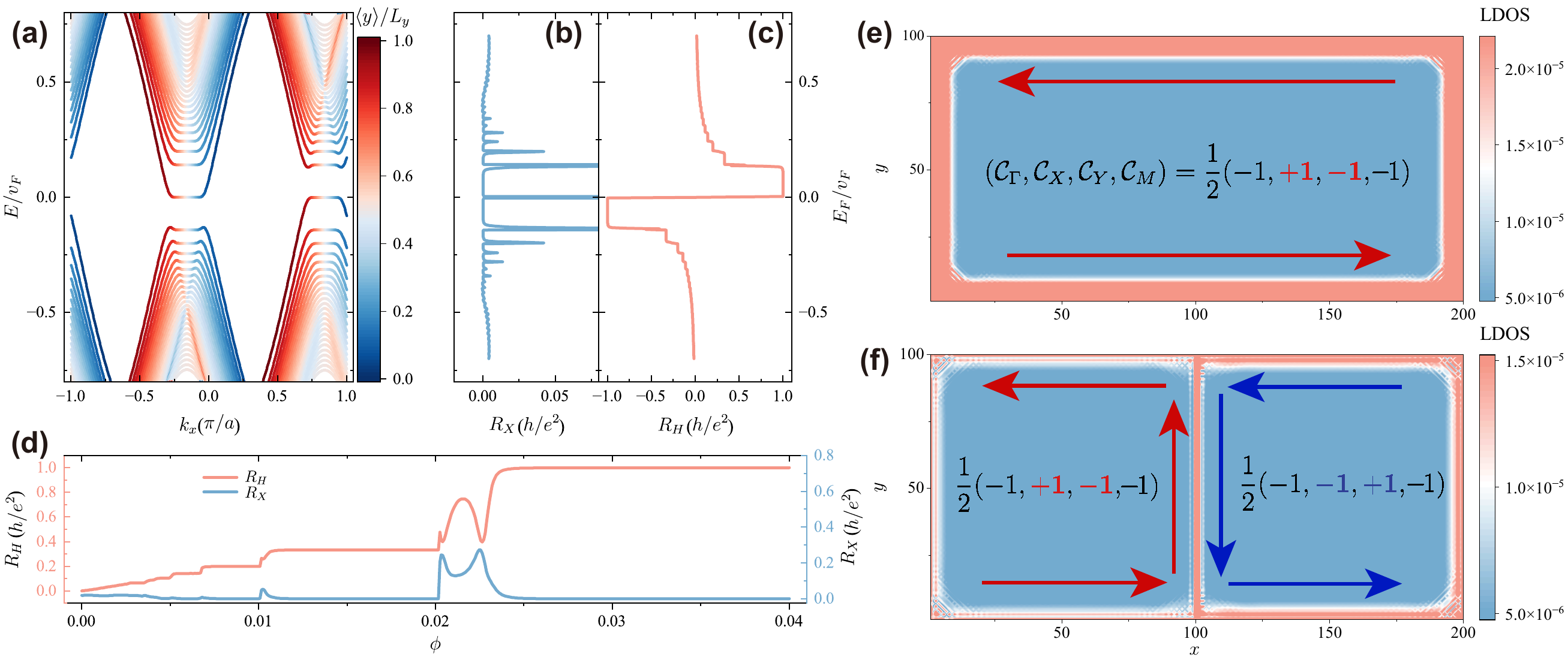}}
	\caption{The behavior of DFAs under magnetic fields.
		(a) The band structure of a DFA nanoribbon, with a width $L_y=100a$ and the magnetic flux $\phi=0.01$ (in the unit of $h/e$).
		(b) and (c) are the longitudinal and Hall resistances of the DFA Hall bar device versus $E_F$, which corresponds to the energy $E$ in (a).
		(d) The longitudinal (blue) and Hall (red) resistances versus $\phi$ with $E_F=0.2v_F$.
		(e,f) are the local density of states of two $100a\times 100a$ coupling DFA samples, with parallel [in (e)] and antiparallel [in (f)] N\'{e}el vectors. $E=0.2$, $\phi=0.03$, and the valley Chern numbers are shown in the insets.
		}
	\label{fig 4}
\end{figure*}

As shown in Fig. \ref{fig 3}(b-d), quantized Hall resistance and vanishing longitudinal resistance indeed emerge while $E_F$ is within the energy gap ($|E_F|<|m|$), as a consequence of the quantized Chern number and chiral edge state transport.
Specifically, as depicted in Fig. \ref{fig 3}(b) and \ref{fig 3}(c), when $m<-|E_F|$ and $m>|E_F|$, the Hall resistance is quantized to $+1$ and $-1$, respectively, accompanied by a longitudinal resistance of $R_X=0$.
Meanwhile, this quantized transport is robust against the choice of $m_{\rm Am}$ and strong disorder (see Sec. S4 and S5 \cite{supp,KVSeO,altermagnet_2022,SCBA,SCBAbreak1,SCBAbreak2}).
When $|m|<|E_F|$, the Hall resistance deviates from the quantized values, and the longitudinal resistance emerges due to contributions from the bulk states.
The hump features near the $m\approx 0$ at $E_F=0$ in Fig. 3(b,d) are related to the size effect (see Sec. S6 \cite{supp,topological_2012,finitegap}).
As illustrated in Fig. \ref{fig 3}(d), by sweeping through $m$ and $E_F$ to calculate $R_H$, we obtain a phase diagram for the DFA, with the QAH insulator phases corresponding to Chern numbers of $-1$ and $+1$ for $m<-|E_F|$ and $m>|E_F|$, respectively.
As shown in Fig. \ref{fig 3}(e), with the help of two identical winding numbers $W=+1$, DFAs exhibit quantized Chern numbers under a Zeeman term, distinctly different from the behavior of $H_{\rm DF}$ and graphene, highlighting the unique role of the altermagnetic mass.

The LLs of electrons exhibiting linear dispersion are particularly interesting.
For instance, graphene possesses symmetrically arranged LLs above and below zero energy \cite{LL_2002,graphene_2005_1,graphene_2005_2}.
The LLs of the electron-type and hole-type of graphene are in touch at $E=0$ symmetrically (see Fig. S5 and Sec. S7 \cite{supp}), protected by the chiral symmetry and Atiyah-Singer index theorem \cite{graphene_2005_1}.
Wilson fermions exhibit asymmetric LLs, with the 0th LL even crossing zero energy, as a signature of chiral-symmetry breaking and parity anomaly \cite{parity_2020}.
Inspired by these observations, we investigate the behavior of 2D DFAs under a perpendicular magnetic field.
The effect of the magnetic field is incorporated into $H_{\rm DFA}$ through the Peierls substitution \cite{Green_2009,Peierls_2008,Peierls_2024}, and the strength of the magnetic field $B$ is evaluated by the flux in each unit cell $\phi=Ba^2$.
As illustrated in Fig. \ref{fig 4}(a), under the influence of the altermagnetic mass $m_{\rm Am}$, the positions of the LLs remain symmetric about $E=0$, similar to the case in graphene.
Figs. \ref{fig 4}(b,c) depicts the quantum Hall transport of chiral edge states between LLs, which leads to the vanished longitudinal and quantized Hall resistance, with $E_F$ corresponding to the energy in Fig. \ref{fig 4} (a).
As shown in Fig. \ref{fig 4}(d), as the magnetic field increases, Hall resistance jumps up one step after another, accompanied by longitudinal resistance peaks.
In transition regions (especially around $\phi=0.02$), the resistances exhibit distinctive lineshapes, as a result of the fine structure of LLs (see Sec. S8 \cite{supp}).
Due to the valley degree of freedom, the Hall resistance is quantized as $\frac{1}{(2n+1)}\frac{h}{e^2}$ \cite{graphene_2005_1,graphene_2005_2}.
However, unlike the QHE in graphene, edge states of DFAs exhibit subtle asymmetry, as a result of chiral-symmetry breaking by the altermagnetic mass.
Specifically, the electron-type (hole-type) LL reaches $E=0$ only around $k_x=0$ ($k_x=\pi/a$).
In other words, the LLs of DFAs possess edge states with complete valley polarization,
which arises from the valley Chern number due to the altermagnetic mass (see details in Sec. S9 \cite{supp}).

It is noteworthy that this valley polarization arises from the influence of the altermagnetic mass term.
When the N\'{e}el vectors of the altermagnets flip, the altermagnetic mass and the valley polarization also undergo a reversal.
The edge states associated with different valleys are well-separated in momentum space, making them hard to couple \cite{classchern_2024}.
Consequently, helical quantum valley Hall edge states can be formed \cite{valley_2009}.
Consider two $100a\times 100a$ square samples possessing DFAs coupled together along the $x$-direction, with a perpendicular magnetic field applied to induce LLs.
When the two samples possess parallel N\'{e}el vectors,, as illustrated in Fig. \ref{fig 4}(e), the perimeter of the $200a\times 100a$ rectangle is surrounded by edge states with valley Chern number $(\mathcal{C}_\Gamma,\mathcal{C}_X,\mathcal{C}_Y,\mathcal{C}_M)=\frac{1}{2}(-1,1,-1,-1)$, with no edge states emerging at the coupling interface.
Conversely, when the N\'{e}el vectors of the two samples are antiparallel, as shown in Fig. \ref{fig 4}(f), despite the identical total Chern numbers on both sides, edge states do emerge at the coupling interface, which are the helical quantum valley Hall edge states arising from the difference of valley Chern number \cite{valley_2009,supp}.
In other words, within DFA systems, the control of helical channels can be achieved by manipulating domain walls of N\'{e}el vectors in altermagnets \cite{neel_2025}.

Viewed from a unifying perspective, the altermagnetic mass directly modifies the Dirac equation and plays a similar conceptual role as the Wilson mass does, with the help of altermagnetic symmetry.
It can be exported to any platform that hosts Dirac-type quasiparticles, including condensed matter physics, high-energy physics, and quantum simulations.
Recent studies that embed altermagnetism into specific models \cite{addb1,addb2,addb3,addb4} illustrate the richness of combining altermagnets and topological physics.
Rather than introducing a mathematically new term, our work shows that the altermagnetic mass provides a distinct and effective route to generating net chirality, which may assist the future design of Chern phases and valley-based functionalities.
Meanwhile, our model is not only applicable to square lattices and $d$-wave altermagnetism, but also has broad possibilities in other systems, such as hexagonal lattices and $g$-wave altermagnetism (see Fig. S8 and Sec. S10 \cite{supp,MnTe1,MnTe2}).
With the continuous discovery of altermagnetic materials in recent years (see Sec. S11 \cite{supp,altermagnet_2022p,altermagnet_2020,RuO22,RuO23,MnTe1,MnTe2,LaMnO3,FeSb,CrSb,new50,CoNbS}), the realization of DFAs appears increasingly promising.

\begin{acknowledgments}
\emph{Acknowledgments}---
P.Y.L. thanks Yu-Chen Zhuang for helpful discussions.
This work was financially supported by
the National Key R and D Program of China (Grant No. 2024YFA1409002),
the National Natural Science Foundation of China (Grants No. 12374034 and Grants No. 124B2069),
the Quantum Science and Technology-National Science and Technology Major Project (Grant No. 2021ZD0302403).
The computational resources are supported by the High-Performance Computing Platform of Peking University.
\end{acknowledgments}

\end{document}


\title{Supplementary Materials for ``The emergence of net chirality in two-dimensional Dirac fermions system with altermagnetic mass''}
\author{Peng-Yi Liu}
\thanks{These authors contributed equally to this work.}
\affiliation{International Center for Quantum Materials, School of Physics,
Peking University, Beijing 100871, China}
\author{Yu-Hao Wan}
\thanks{These authors contributed equally to this work.}
\affiliation{International Center for Quantum Materials, School of Physics,
Peking University, Beijing 100871, China}
\author{Qing-Feng Sun}
\email[Corresponding author: ]{sunqf@pku.edu.cn}
\affiliation{International Center for Quantum Materials, School of Physics,
Peking University, Beijing 100871, China}
\affiliation{Hefei National Laboratory, Hefei 230088, China}
\date{\today }

\maketitle
\tableofcontents
\clearpage
\section{Winding number of spin texture and chirality of Dirac fermions}
Based on the fact that the spin texture $\bm{S=\langle \sigma \rangle}$ of the Dirac fermion (DF) in the momentum space has a vortex structure [as shown in Fig. 1(b) in the main text], the winding number can be calculated as follows.

Around a Dirac point, the spin expectation of the occupied state can be expressed as $\bm{S}=[\cos \theta(\varphi), \sin \theta(\varphi),0]$, where $\varphi$ is the azimuthal angle in the two-dimensional Brillouin zone.
Then, the winding number is given by Eq. (\ref{eq:winding}) \cite{winding_2007,winding_2021}.
\begin{equation}
W=\int_0^{2\pi}{\rm{d}}\varphi \frac{1}{2\pi} \frac{\partial \theta}{\partial \varphi}=\frac{1}{2\pi}\theta(\varphi)\big|_{\varphi=0}^{\varphi=2\pi} .\label{eq:winding}
\end{equation}

For the Dirac cones of $H_{\rm DF}$ at $\Gamma$ and $M$ points, we can take $\theta(\varphi)=\varphi+n\pi$ ($n$ is an integer), and the winding number is $W=+1$.
That is to say, clockwise around the Dirac point, the spin texture also rotates clockwise.
In contrast, for $X$ and $Y$ points, we can take $\theta(\varphi)=-\varphi+n\pi$, and the winding number is $W=-1$.
Using winding number, we get a way to capture the chiral property of a DF.
As mentioned in the main text, the chiralities of $\Gamma$, $X$, $Y$, and $M$ points are $+$, $-$, $-$, and $+$, respectively.

For DFs with altermagnetic mass (DFAs), the strict chiral symmetry is broken, and as a result, the spin texture acquires an out-of-plane component $\bm{S}=[\cos \theta(\varphi), \sin \theta(\varphi),S_z(k_x,k_y)]$, where $S_z$ is a scalar field in the Brillouin zone.
However, the altermagnetic mass depends on the square of the momentum, and the symmetry is approximately restored at positions close enough to the Dirac point.
Meanwhile, it can be checked by Eq. (\ref{eq:winding}) that the winding numbers near the Dirac points of $H_{\rm DFA}$ $\Gamma$ and $M$ remain unchanged $W=+1$.
This inspires us that, in 2D lattice systems, DFAs have properties similar to two DFs with the same chirality.

\section{Lattice Hamiltonian}
The tight-binding Hamiltonian of the lattice DF, which we used to calculate the band structure of the nanoribbon and transport, is connected to the continuous one through a Fourier transformation.
Here, the discrete lattice constant $a$ has been set to the unit length $a=1$.
\begin{equation}
    \begin{aligned}
    H_{\rm{DF}}&=\sum_{\bm{k}}{v_F}c_{\bm{k}}^\dagger (\sin k_x \sigma_x+\sin k_y \sigma_y) c_{\bm{k}}\\
    &=\frac{1}{L_x L_y} \sum_{x,y}\sum_{x',y'}\sum_{k_x,k_y}{v_F}c_{x,y}^\dagger e^{{\rm{i}}k_xx+{\rm{i}}k_yy} ({\rm{sin}}k_x \sigma_x+{\rm{sin}}k_y \sigma_y) c_{x',y'}e^{-{\rm{i}}k_xx'-{\rm{i}}k_yy'}\\
    &=\frac{1}{L_x L_y} \sum_{x,y}\sum_{x',y'}\sum_{k_x,k_y}{v_F}c_{x,y}^\dagger \left\{\frac{\sigma_x}{2{\rm{i}}}\left[e^{{\rm{i}}k_x(x-x'+1)}-e^{{\rm{i}}k_x(x-x'-1)}\right]e^{{\rm{i}}k_y(y-y')} \right.\\
   &\hspace{42mm} \left.+\frac{\sigma_y}{2{\rm{i}}}\left[e^{{\rm{i}}k_y(y-y'+1)}-e^{{\rm{i}}k_y(y-y'-1)}\right]e^{{\rm{i}}k_x(x-x')}\right\}
    c_{x',y'}\\
    &=\sum_{x,y}
     c_{x,y}^\dagger \frac{{v_F}}{2{\rm{i}}}\sigma_x c_{x+1,y} + c_{x,y}^\dagger \frac{{v_F}}{2{\rm{i}}}\sigma_y c_{x,y+1}+h.c.,
    \end{aligned}\label{eq:HDF}
\end{equation}
where $c_{x,y}^\dagger=[c_{x,y;\uparrow}^\dagger,c_{x,y;\downarrow}^\dagger]$ is the creation operator of electrons at site $(x,y)$. $\uparrow$ and $\downarrow$ represent the spin degree of freedom.
$L_x$ and $L_y$ are the width in the $x$ and $y$-directions.
Similarly, the discrete version of the Hamiltonian of the DF with altermagnetic mass (DFA) is given as:
\begin{equation}
    \begin{aligned}
    H_{\rm{DFA}}&=\sum_{\bm{k}}c_{\bm{k}}^\dagger \left[{v_F}\sin k_x \sigma_x+{v_F}\sin k_y \sigma_y + m_{\rm Am}(\cos k_x -\cos k_y)\sigma_z\right] c_{\bm{k}}\\
    &=\frac{1}{L_x L_y} \sum_{x,y}\sum_{x',y'}\sum_{k_x,k_y}c_{x,y}^\dagger e^{{\rm{i}}k_xx+{\rm{i}}k_yy} \left[{v_F}\sin k_x \sigma_x+{v_F}\sin k_y \sigma_y + m_{\rm Am}(\cos k_x -\cos k_y)\sigma_z\right] c_{x',y'}e^{-{\rm{i}}k_xx'-{\rm{i}}k_yy'}\\
    &=\frac{1}{L_x L_y} \sum_{x,y}\sum_{x',y'}\sum_{k_x,k_y}c_{x,y}^\dagger
    \left\{\frac{{v_F}\sigma_x}{2{\rm{i}}}\left[e^{{\rm{i}}k_x(x-x'+1)}-e^{{\rm{i}}k_x(x-x'-1)}\right]e^{{\rm{i}}k_y(y-y')} \right.\\
        &\hspace{40mm}
    +\frac{{v_F}\sigma_y}{2{\rm{i}}}\left[e^{{\rm{i}}k_y(y-y'+1)}-e^{{\rm{i}}k_y(y-y'-1)}\right]e^{{\rm{i}}k_x(x-x')}\\
    &\hspace{40mm}
    +\frac{m_{\rm Am}\sigma_z}{2}\left[e^{{\rm{i}}k_x(x-x'+1)}+e^{{\rm{i}}k_x(x-x'-1)}\right] e^{{\rm{i}}k_y(y-y')} \\
    &\hspace{40mm}  \left.
    -\frac{m_{\rm Am}\sigma_z}{2}\left[e^{{\rm{i}}k_y(y-y'+1)}+e^{{\rm{i}}k_y(y-y'-1)}\right] e^{{\rm{i}}k_x(x-x')}\right\}
    c_{x',y'}\\
    &=\sum_{x,y}
     c_{x,y}^\dagger \left(\frac{{v_F}}{2{\rm{i}}}\sigma_x+\frac{m_{\rm Am}}{2}\sigma_z \right)c_{x+1,y} + c_{x,y}^\dagger \left(\frac{{v_F}}{2{\rm{i}}}\sigma_y-\frac{m_{\rm Am}}{2}\sigma_z\right) c_{x,y+1}+h.c..
    \end{aligned}\label{eq:HDFA}
\end{equation}

In the main text, we use the lattice Hamiltonian to construct the band structure of the nanoribbons and calculate the electron transport.

\begin{figure}[htbp]
 \centering
 \includegraphics[width=1\columnwidth]{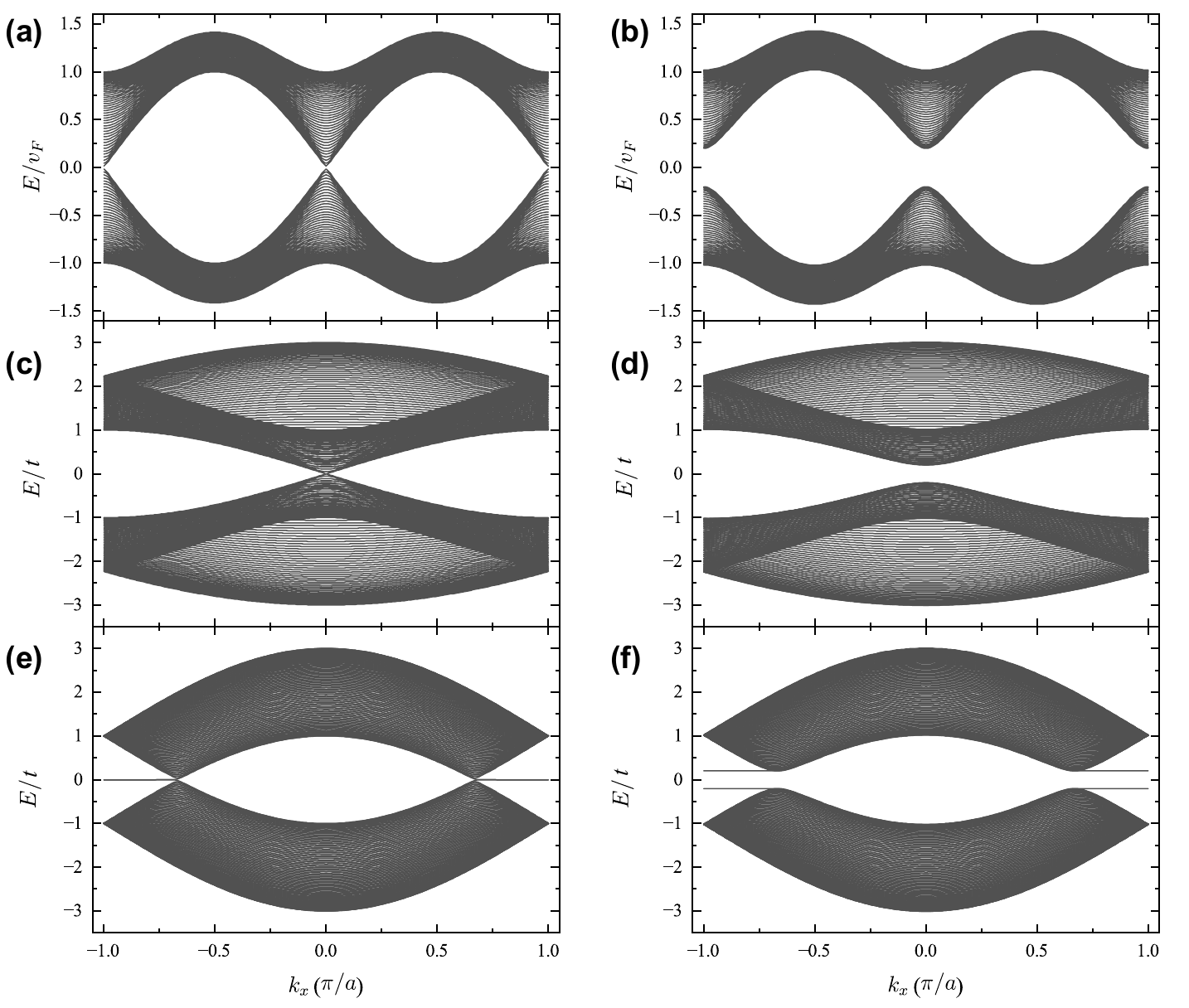}
 \caption{Band structure of gapless (a, c, e) and gapped (b, d, f) DFs without altermagnetic mass.
 (a) The energy bands of a nanoribbon, whose Hamiltonian is given by Eq. ({\ref{eq:HDF}}).
 (b) The energy bands of a nanoribbon, whose Hamiltonian is $H=H_{\rm{DF}}+m\sigma_z$ with $m=0.2 v_F$.
 (c) and (d) are the energy bands (in the unit of the nearest-neighbor hopping $t$) of armchair graphene nanoribbons without and with a Zeeman term $m\sigma_z$, respectively, and $m=0.2t$.
 (e) and (f) are the energy bands (in the unit of the nearest-neighbor hopping $t$) of zigzag graphene nanoribbons without and with a Zeeman term $m\sigma_z$, respectively.
 The unit of energy is $v_F$ for $H_{\rm DF}$ and is the nearest-neighbor hopping for graphene.
 The width of these nanoribbons in the $y$ direction are $L_y=100a$ (a,b), $50a$ (c,d), and $\frac{(3\times 50 -1)a}{\sqrt{3}}$ (e,f), respectively, where $a$ is the lattice constant for each system.
 }
 \label{FIGS1}
\end{figure}

\section{Band structure of DFs without altermagnetic mass}

In the main text, we find that the nanoribbon of gapped DFA $H=H_{\rm{DFA}}+m\sigma_z$ has chiral edge states in the gap.
To test that this is a special property of DFA, as shown in Fig. \ref{FIGS1}, we examined the band structure of $H_{\rm DF}$ [Fig. \ref{FIGS1}(b)], armchair graphene [Fig. \ref{FIGS1}(d)], and zigzag graphene [Fig. \ref{FIGS1}(f)] nanoribbons after they were opened by the Zeeman term $m\sigma_z$.
Before the introduction of $m\sigma_z$, as shown in Fig. \ref{FIGS1}(a,c,e),
their upper and lower bands can touch, showing gapless Dirac behavior.
However, when the Zeeman term turns on, the energy gaps are opened, and no edge state appears in the energy gaps.
It is not difficult to verify that the Chern number within these gaps is 0 because the Berry curvatures contributed by the Dirac cones with different chirality cancel each other out.

\section{The influence of the amplitude of the alternating magnetic mass.}

In the main text, we use the value \( m_{\mathrm{Am}} = 0.4v_F \), which is a representative choice, not fine-tuned.
Our key conclusions do not rely on this specific value.
More precisely, in the Hamiltonian \( H_{\rm DFA}+m \sigma_z\equiv \bm{d}\cdot \bm{\sigma} \), the topological phase boundaries are determined solely by gap closing and reopening at four high-symmetry points in the Brillouin zone.
The resulting analytic condition is
\begin{equation}
d_z(\Gamma)d_z(X)d_z(Y)d_z(M)=m^2(m + 2 m_{\mathrm{Am}})(m - 2 m_{\mathrm{Am}}) = 0,
\end{equation}
implying that the system resides in a QAH phase with \(|C| = 1\) whenever \( 0 < |m| < 2|m_{\mathrm{Am}}| \).
Thus, the precise value of \( m_{\mathrm{Am}} \) does not qualitatively affect the topology.

\begin{figure}
	\includegraphics[width=0.8\columnwidth]{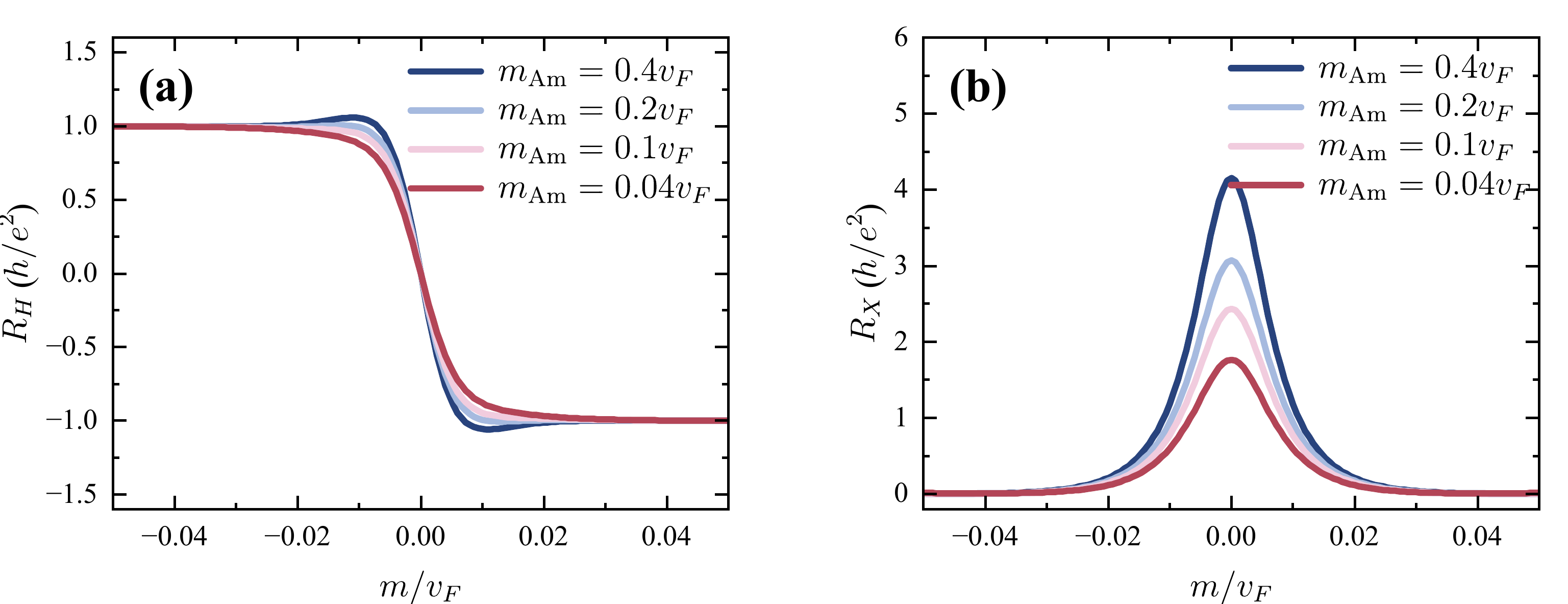}
			\caption{\label{FIGS2}
				\baselineskip 4pt
	The altermagnetic mass $m_{\rm Am}$ dependence of QAH transport.
				(a) and (b) are the repetition of the curves ``$E_F=0$" in Fig. 3(b,c) ($m_{\rm Am}=0.4v_F$) in the main text but with smaller altermagnetic mass $m_{\rm Am}$ for the QAH transport calculation.
				}
\end{figure}

To place these model parameters into a realistic context, we refer to recent experimental and first-principles studies on altermagnetic materials that share the same symmetry structure relevant to our model.
For example, in the recently identified altermagnet $\rm KV_2Se_2O$, combined ARPES measurements and density-functional theory (DFT) calculations reveal that the bands at $\Gamma$ and $M$ remain spin-degenerate, while those at the $X$ and $Y$ points exhibit momentum-dependent spin splittings of approximately 1.6 eV \cite{KVSeO}.
Likewise, previous DFT studies of $\rm RuO_2$ report splittings of order 1 eV at $X/Y$, with degenerate states at $\Gamma/M$ \cite{RuO2}.
Although our model is intentionally minimal, it incorporates the same alternation of degeneracy and splitting across the $\Gamma/M$ and $X/Y$ points.
This allows us to extract a rough estimate of the altermagnetic mass term in our notation:
${m_{\mathrm{Am}}} \approx 0.5\sim0.8~\text{eV}$, 
corresponding to the experimentally and computationally observed energy scales.

These material-based estimates also indicate that the altermagnetic mass lies notably below the full bandwidth, suggesting that in realistic systems, both \( m_{\mathrm{Am}} \) and \( m \) are much smaller than the overall electronic energy scale.
To address this, we repeat the transport simulations in Fig. 3(b,c) using smaller values of \( m_{\mathrm{Am}} \) and \( m \), while keeping the Fermi energy fixed at \( E_F = 0 \).
As shown in Fig. \ref{FIGS2}, the quantization of the Hall resistance and vanishing of the longitudinal resistance remain robust, and the curves are nearly unchanged.
Slight deviations occur only when $m\approx0$, where the system approaches the gapless point and finite-size effects become non-negligible.

\section{Robustness of the net-chiral transport}

For experimental observations, an interesting and important question is whether the observable effects of the proposed chirality mechanism are robust against disorder, as disorder inevitably exists in the real system.

We first analyze the effect of disorder on the net-chiral Dirac points by computing the disorder-averaged self-energy using a standard disorder Green's function formalism under the self-consistent Born approximation (SCBA) \cite{SCBA}.
This provides analytical insight into how disorder modifies the low-energy Dirac structure.
The disordered potential enters the Hamiltonian as an onsite term:
\begin{equation}
H_{\rm Born}^{\text{disorder}} = \sum_{x,y}  c^\dagger_{x,y} U(x,y) c_{x,y}, \quad U(x,y) = U_\uparrow(x,y)|x,y;\uparrow\rangle\langle x,y;\uparrow|+U_\downarrow(x,y)|x,y;\downarrow\rangle\langle x,y;\downarrow|.
\end{equation}
Here, $U_\uparrow(x,y)$ and $U_\downarrow(x,y)$ are independent and uniformly distributed in the region $[-W_B/2,W_B/2]$.
For the spin-uncorrelated disorder, $\langle U_\uparrow U_\uparrow \rangle=\langle U_\downarrow U_\downarrow \rangle=W_B^2/12$ and $\langle U_\uparrow U_\downarrow \rangle=0$, where $\langle \dots \rangle$ represents the disorder averaging.
Under these conditions, the self-consistent Born self-energy reads:
\begin{equation}
\Sigma^{r/a}(\varepsilon) = \sum_{\mathbf{k}} \langle \begin{bmatrix}
U_\uparrow & 0 \\
0 & U_\downarrow
\end{bmatrix} G^{r/a}(\varepsilon,\bm{k}) \begin{bmatrix}
U_\uparrow & 0 \\
0 & U_\downarrow
\end{bmatrix} \rangle,
\end{equation}
where $r$ and $a$ label retarded and advanced self-energy or Green's function, respectively.
For the lowest-order approximation, $G^{r/a}(\varepsilon,\bm{k})\rightarrow G_{(0)}^{r/a}(\varepsilon,\bm{k})=\left[ (\varepsilon \pm {\rm i}0^+)- H_{DFA}(\bm{k}) \right]^{-1}$.
And the self-energy under the zero-order self-consistent reads:
\begin{equation}
\begin{aligned}
\Sigma_{(0)}^{r/a}(\varepsilon) &= \sum_{\mathbf{k}} \frac{W_B^2}{12} \frac{\begin{bmatrix}
{\varepsilon+m_{\rm Am}[\cos (k_x)-\cos(k_y)]} & 0 \\
0 & {\varepsilon-m_{\rm Am}[\cos (k_x)-\cos(k_y)]}
\end{bmatrix}}{(\varepsilon \pm {\rm i}0^+)^2 - v_F^2[\sin(k_x)^2+\sin(k_y)^2]-m_{\rm Am}^2 [\cos (k_x)-\cos(k_y)]^2}\\
&=\sum_{\mathbf{k}} \frac{W_B^2}{12} \frac{\varepsilon}{(\varepsilon \pm {\rm i}0^+)^2 - v_F^2[\sin(k_x)^2+\sin(k_y)^2]-m_{\rm Am}^2 [\cos (k_x)-\cos(k_y)]^2} \sigma_0\\
&\propto \sigma_0.
\end{aligned}
\end{equation}
Here, the lattice constant has been set to 1.
During the integration process, the altermagnetic mass term $m_{\rm Am}[\cos (k_x)-\cos(k_y)]$ in the numerator is eliminated due to its parity, so the zeroth-order self-energy is proportional to $\sigma_0$.
According to the self-consistent process:
\begin{equation}
\begin{aligned}
\Sigma_{(n)}^{r/a}(\varepsilon) &= \langle \begin{bmatrix}
U_\uparrow & 0 \\
0 & U_\downarrow
\end{bmatrix} G_{(n)}^{r/a}(\varepsilon,\bm{k}) \begin{bmatrix}
U_\uparrow & 0 \\
0 & U_\downarrow
\end{bmatrix} \rangle \propto  \sigma_0 \\
G_{(n)}^{r/a} &= \left[ (\varepsilon \pm {\rm i}0^+) - H_{\rm DFA}(\bm{k}) -\Sigma_{(n-1)}^{r/a} \right]^{-1}.
\end{aligned}
\end{equation}
The disorder-averaged self-energy at any order is always proportional to the identity matrix $\sigma_0$.
Therefore, disorders will not destroy the net-chiral properties, at the level of the SCBA \cite{SCBA}.
This remains true for both the low-energy and lattice-regularized versions of our model.
We also emphasize that the analytical discussion here is restricted to the zero-magnetic-field case to avoid the breakdown of SCBA upon the formation of well-separated Landau levels and edge-state structures \cite{SCBAbreak1,SCBAbreak2}.
As shown in Fig. 4 of the main text, under strong magnetic fields, the formation of Landau levels leads to quantized Chern numbers and topologically protected chiral edge states, which are known to be robust against disorder.
\\

\begin{figure}
	\includegraphics[width=\columnwidth]{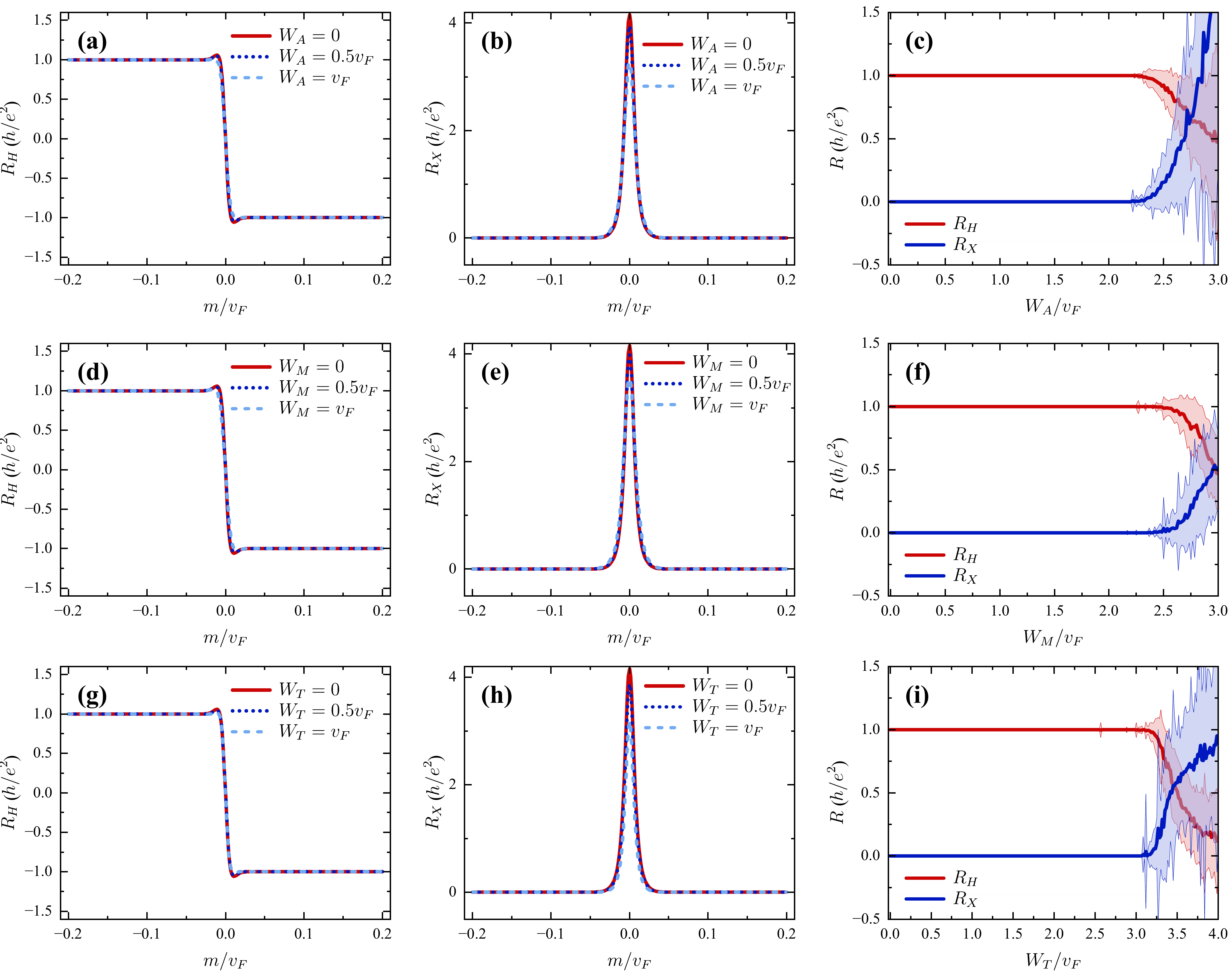}
			\caption{\label{FIGS3}
				\baselineskip 4pt {
				Numerical results of on-site disorder (a-c), magnetic disorder (d-f), and hopping disorder (g-i), which verify the robustness of net-chirality.
				The curves in (a, d, g) are directly compared with the results ``$E_F=0$'' shown in Fig. 3(b) in the main text.
				The curves in (b), (e), and (h) are directly compared with the results ``$E_F=0$'' shown in Fig. 3(c) in the main text.
				(c, f, i) show the variation of resistance with the disorder strength.
				$E_F=0$ and $m=-0.2 v_F$ are fixed.
				The shaded areas represent the standard deviation of the corresponding resistance across all disorder configurations.
				Each situation was calculated using 200 sets of random disorder configurations.		}		
				}
\end{figure}

Having established analytically that disorder cannot destroy the net-chiral Dirac points on the SCBA level, we now turn to a more in-depth numerical study that goes beyond the SCBA and includes more types of disorder.
Two practical issues motivate our numerical strategy:
\begin{itemize}
\item Once translational symmetry is broken by disorder, momentum $\bm{k}$ is no longer a good quantum number, so Dirac points in momentum space can at best be defined approximately.
Transport quantities, by contrast, remain well defined.
\item When a trivial mass term $m\sigma_z$ opens a bulk gap, the net-chiral characteristic directly leads to QAH transport, and QAH transport is a direct manifestation of net chirality.
The persistence (or breakdown) of this QAH plateau under disorder therefore provides a direct and faithful measure of how robust the underlying net-chiral topology is.
\end{itemize}

Guided by these considerations, we perform calculations of three types of disorders and repeat the curves in Fig. 3(b,c) with $E_F=0$ for direct comparison:

\begin{itemize}
  \item On-site (Anderson) potential disorder, $H_{\rm on-site}^{\rm disorder}=\sum_{x,y}\delta E_{x,y} c_{x,y}^\dagger c_{x,y}$, with $\delta E_{x,y}$ uniformly randomly distributed in $[-W_A/2,W_A/2]$, as shown in Fig. \ref{FIGS3}(a-c);
  \item Zeeman disorder, simulating spatial fluctuations of the magnetic order, $H_{\rm Zeeman}^{\rm disorder}=\sum_{x,y}\delta M_{x,y} c_{x,y}^\dagger \sigma_z c_{x,y}$, with $\delta M_{x,y}$ uniformly randomly distributed in $[-W_M/2,W_M/2]$, as shown in Fig. \ref{FIGS3}(d-f);
  \item Hopping (bond) disorder, representing lattice distortions, $H_{\rm hopping}^{\rm disorder}=\sum_{x,y}\left[\frac{\delta v_{x,y}^x}{2 {\rm i}} c_{x,y}^\dagger \sigma_x c_{x+1,y}  +  \frac{\delta v_{x,y}^y}{2 {\rm i}} c_{x,y}^\dagger \sigma_y c_{x,y+1} \right.+\left. {\rm H.c.}\right]$, with amplitudes $\delta v_{x,y}^x$ and $\delta v_{x,y}^y$ uniformly randomly distributed in $[-W_T/2,W_T/2]$, as shown in Fig. \ref{FIGS3}(g-i).
\end{itemize}

For each disorder strength, we averaged over 200 random configurations, with the results summarized in Fig. \ref{FIGS3}.
The quantized Hall plateau and the lineshape persist up to disorder strength of order $W_A,W_N,W_T\gg |m|$, where $m$ is the trivial mass and bulk gap in the clean case.
Even when the disorder strength is close to the bandwidth, i.e., $W_A,W_N,W_T= v_F$, the quantized curve shape remains almost unchanged.
Only when the disorder strengths become much larger than the full bandwidth does the plateau begin to break down.
These calculations demonstrate that the net-chiral topology caused by the altermagnetic mass is remarkably robust against realistic imperfections.

\section{Size effect of the QAH transport}

\begin{figure}
	\includegraphics[width=0.7\columnwidth]{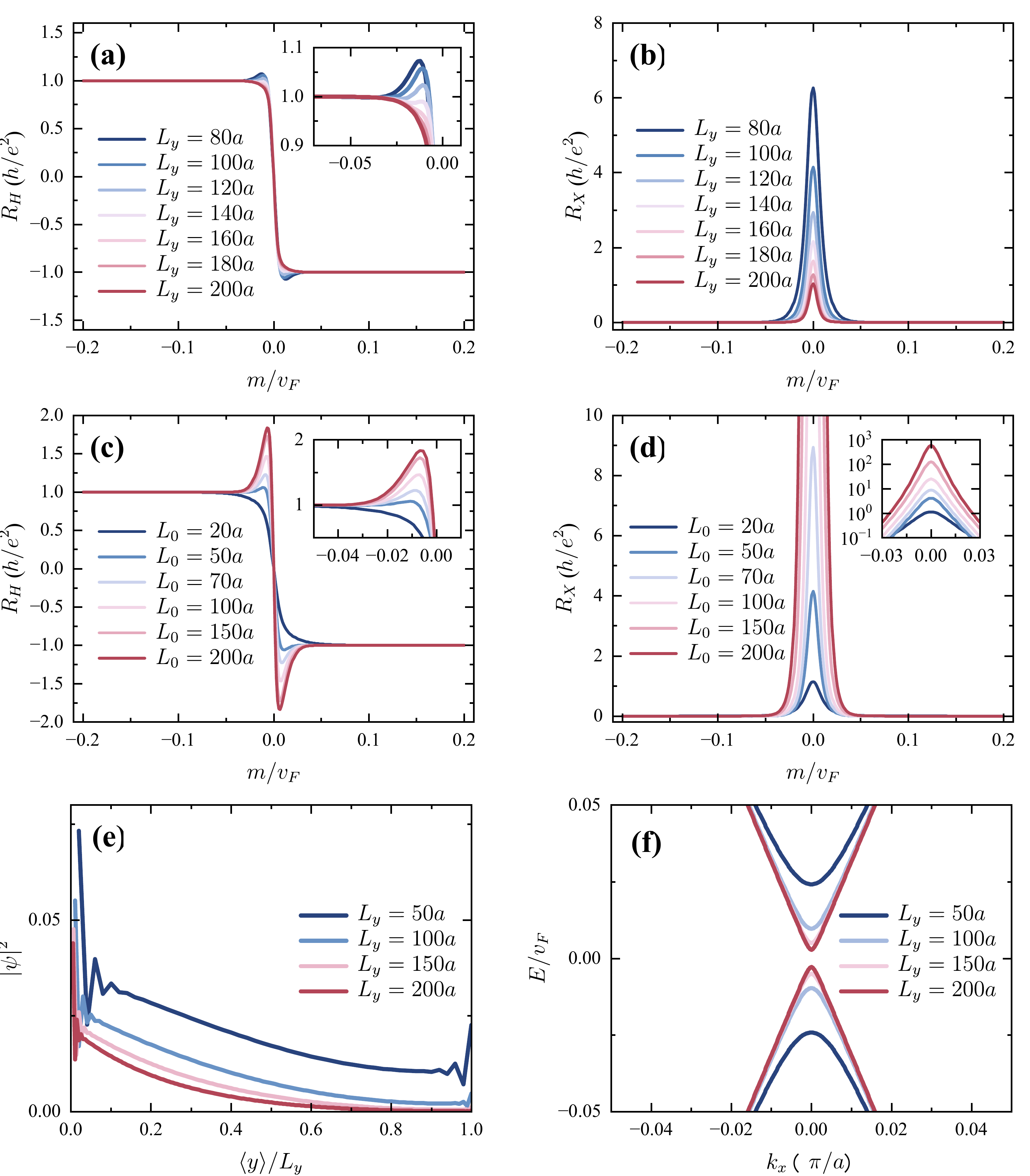}
			\caption{\label{FIGS4}
				\baselineskip 4pt
{	The size dependence of the QAH effect.
				(a,b) Repetition of the $E_F=0$ curves of Fig. 3(b,c) ($L_y=100a$) in the main text under different $L_y$ values.
				(c,d) Repetition of the $E_F=0$ curves of Fig. 3(b,c) ($L_0=50a$) in the main text under different $L_0$ values.
The insets in (a,c,d) are local magnifications of the corresponding main figures.
				(e) Modular squares of the wavefunctions of positive-energy edge states at $k=-0.005\pi/a$ under different $L_y$, with a small trivial mass $m=-0.01$.
				(f) The energy band of DFAs nanoribbons in the $x$-direction around $k_x=0$.
				Only the outermost states are shown and $m=-0.01$.
				A small energy gap opened by the edge-state coupling, with different $L_y$.
			}	}
\end{figure}

In Fig. 3(b-d) of the main text, at the positions where $E_F = 0$ and $m \approx 0$, $R_H$ bulges out as the phase transition approaches, accompanied by a significant peak in $R_X$.
Next, we will see that this is caused by the size effect, by repeating the \( E_F = 0 \) calculations of Fig. 3(b,c) for different Hall-bar dimensions.
\begin{itemize}
\item Variation of the length of the transverse (\( y \)) direction:
Keeping \( L_x = L_0 = 50a \), we varied \( L_y \) from \( 80a \) to \( 200a \) [Fig. \ref{FIGS4}(a,b)].
The quantized values \( R_{H}=h/e^{2} \) and \( R_{X}=0 \) are maintained almost everywhere, but the anomaly near \( m = 0 \) (a slight bump in \( R_{H} \) and a peak in \( R_{X} \)).
As \( L_y \) increases, this anomalous behavior becomes progressively weaker, with the reduction in the width and amplitude of the anomalous area.

\item Variation of the length of the longitudinal (\( x \)) direction:
Next, we fixed \( L_y = 100a \) and $L_x=50a$ (the same as that in the main text) and enlarged \( L_0 \) from \( 20a \) to \( 200a \) [Fig. \ref{FIGS4}(c,d)].
The sharp feature grows with \( L_0 \).
Particularly, \( R_{X} \) increases nearly exponentially (see the inset of Fig. \ref{FIGS4}(d)) and \( R_{H} \) shows a pronounced deviation.
\end{itemize}

When \( |m| \) is small, the decay length of the QAH edge modes is long and becomes comparable to or larger than \( L_y/2 \).
Opposite edges therefore hybridize and open a finite-size gap, as illustrated in Fig. \ref{FIGS4}(e,f), which has been discussed in previous studies about the Wilson mass  \cite{topological_2012,finitegap}.
A smaller $m_{\rm Am}$ or a larger $L_y$ suppresses this gap and hence the sharp feature, whereas increasing \( L_0 \) extends the insulating segment along which back-scattering occurs, producing the exponential rise in \( R_{X} \) and the concurrent deviation in \( R_{H} \).
These trends are fully consistent with the numerical results in Fig. \ref{FIGS4}(a-d).

\section{Landau levels of DFs without altermagnetic mass}
\begin{figure}[htbp]
 \centering
 \includegraphics[width=0.7\columnwidth]{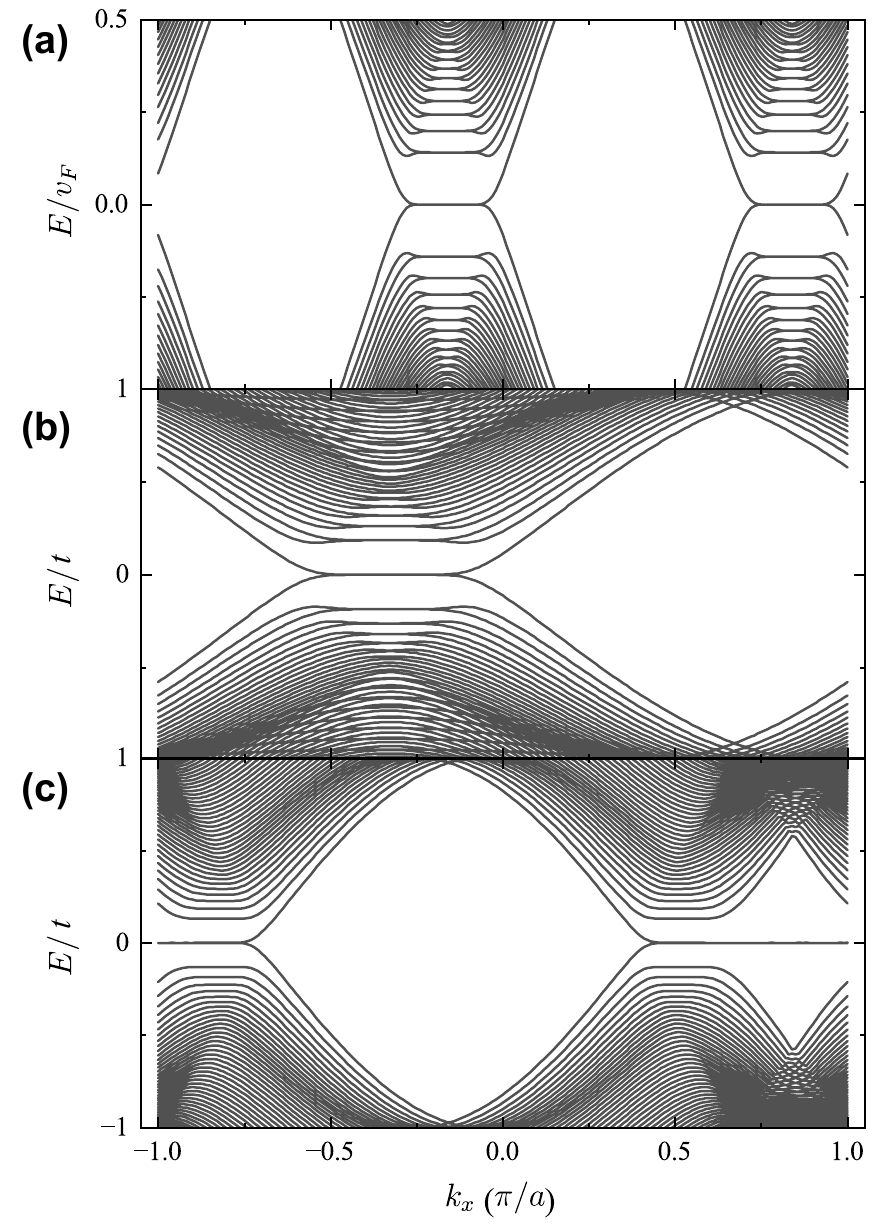}
 \caption{Landau levels of a $H_{\rm DF}$ nanoribbon (a), an armchair graphene nanoribbon (b), and a zigzag graphene nanoribbon (c), with the magnetic flux in each unit cell $\phi=0.01\phi_0$.
 The widths in the $y$ direction and the energy units are the same as those in Fig. \ref{FIGS1}.
 }
 \label{FIGS5}
\end{figure}

In the main text, we find that the chiral edge states of a DFA nanoribbon under a perpendicular magnetic field show a unique asymmetry and valley polarization [see Fig.4(a) in the main text].
In Fig. \ref{FIGS5}, we show the Landau levels of three types of nanoribbons with Dirac cones under magnetic fields, without the presence of an altermagnetic mass.
All Landau levels and chiral edge states show symmetry with respect to $E=0$, and no valley polarization appears.

\section{The fine structure of the Landau level}

\begin{figure}
	\includegraphics[width=0.8\columnwidth]{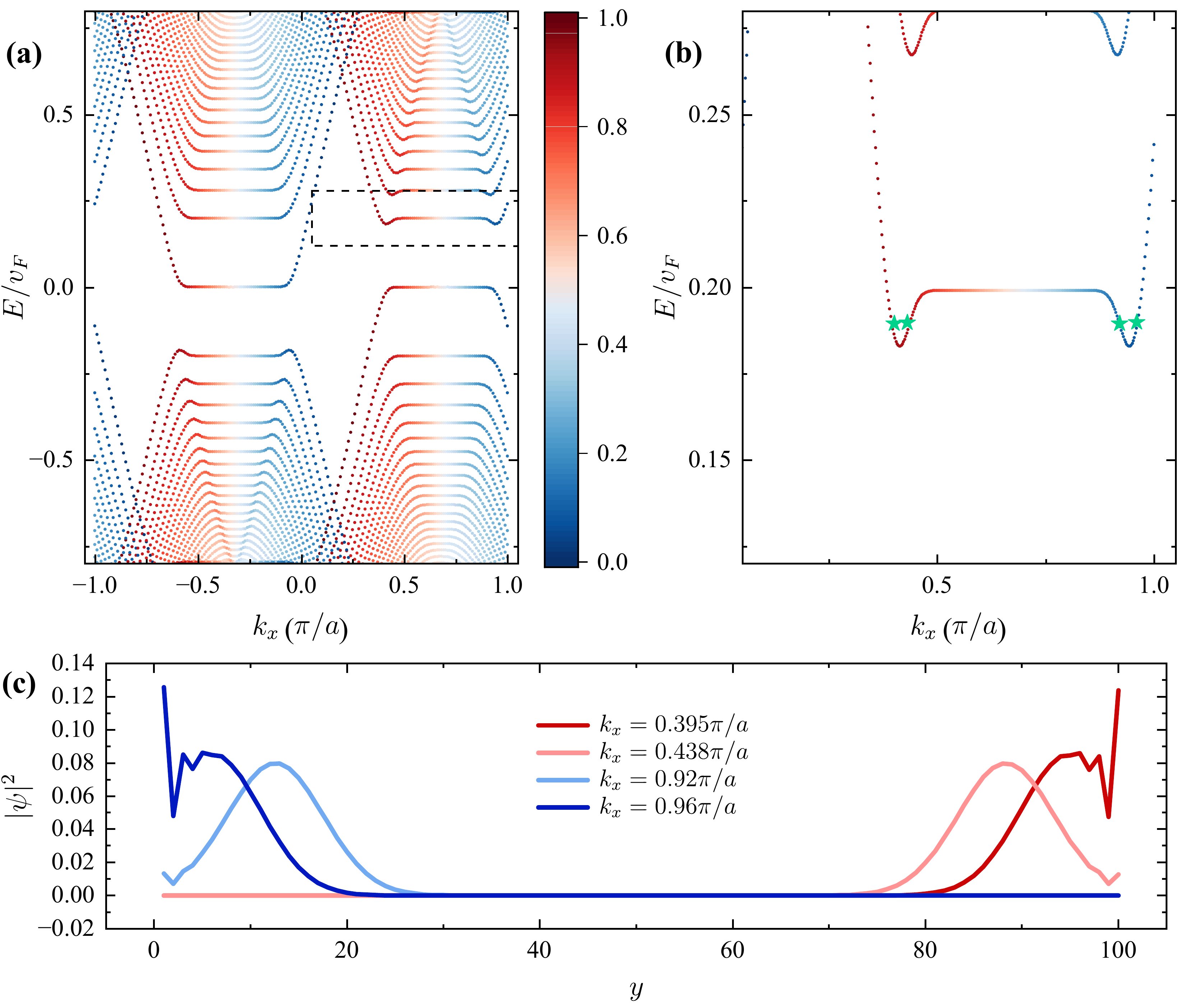}
			\caption{\label{FIGS6}
				\baselineskip 4pt {
				(a) The band structure of Landau levels similar to Fig. 4(a) of the main text, with $\phi$=0.02. The colorbar shows the position of each state $\langle y \rangle /L_y$.
				(b) is a magnification of the dashed box of (a).
				(c) Modular squares of the wavefunctions in the y-direction of four states marked by four green stars in (b).}
				}
\end{figure}

As shown in Fig. 4(d) of the main text, around $\phi = 0.02$, both the Hall and longitudinal resistances show interesting anomalous lineshapes.
We attribute this anomalous lineshape to the unique structure of Landau levels in the presence of the altermagnetic mass.
Specifically, the first Landau level exhibits a non-monotonic bending near the edge toward the zeroth Landau level, as we show in Fig. \ref{FIGS6}(a,b).
This creates a narrow energy window where counter-propagating edge modes coexist along the same edge, which originates from the combined effect between valleys shown in Fig. \ref{FIGS7}.
Fig. \ref{FIGS6}(b) zooms in on the dashed boxed region in Fig. \ref{FIGS6}(a), where the energy window contains these counter-propagating states.
To confirm their nature, we plot the corresponding wavefunctions $|\psi|^2$
in Fig. \ref{FIGS6}(c).
The four states marked by green stars in panel (b) form two pairs, each localized at opposite edges of the sample; within each edge, the two states exhibit opposite group velocities (i.e., counter-propagating), as indicated by their positive and negative band slopes.

This localized coexistence of forward- and backward-moving edge states leads to enhanced backscattering, and thereby causes the oscillatory deviation from quantized values observed in Fig. 4(d) around $\phi = 0.02$.
Unlike in conventional QH systems, where edge states are unidirectional, here the altermagnetic mass leads to subtle structure in the spectrum and transport.

\section{The origin of the asymmetry of the Landau levels}

\begin{figure}
	\includegraphics[width=0.9\columnwidth]{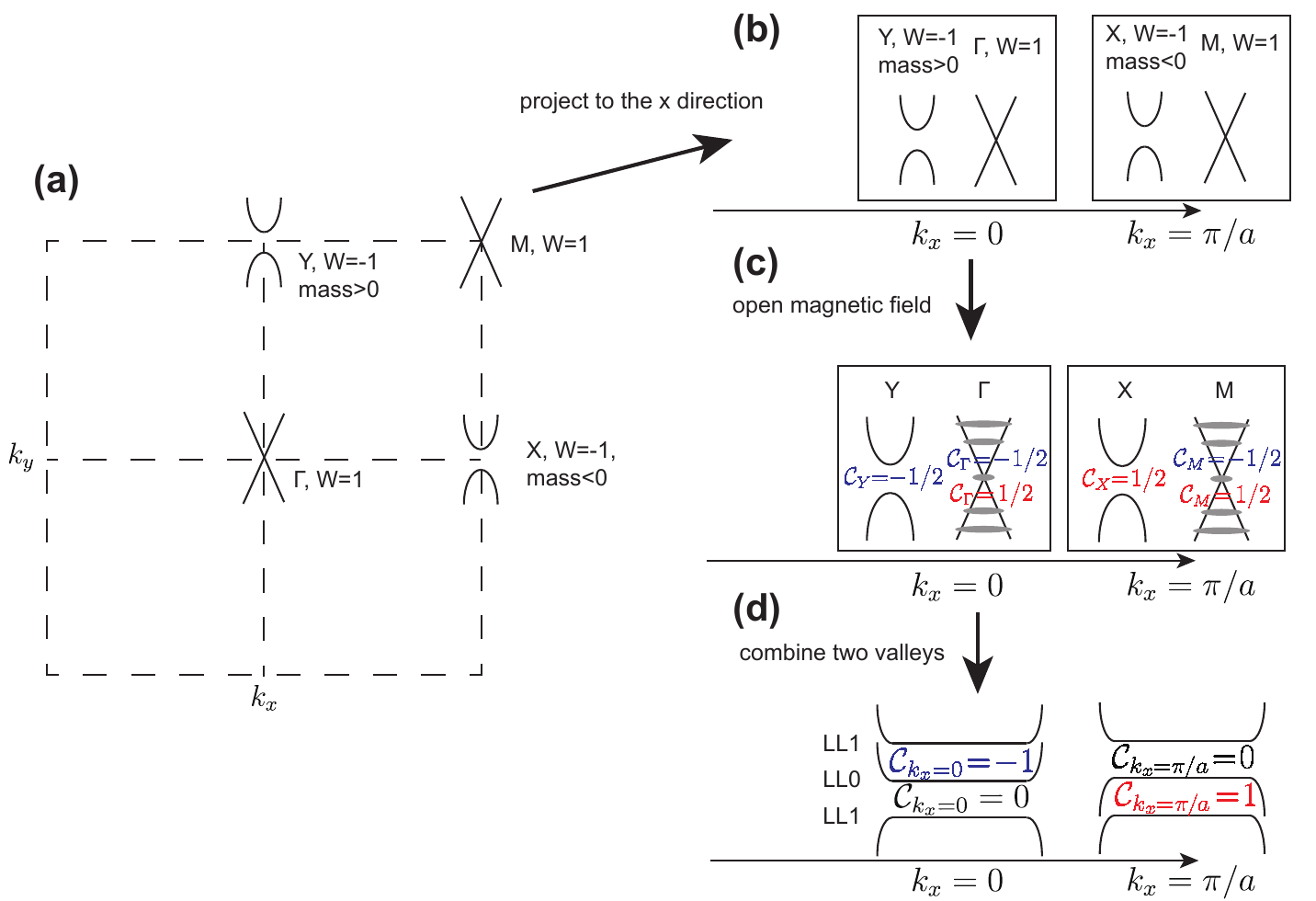}
			\caption{\label{FIGS7}
				\baselineskip 4pt {
				(a) The first Brillouin zone contains four inequivalent valleys ($\Gamma$, $X$, $Y$, and $M$), with winding numbers $W=\pm 1$.
				(b) Project the Brillouin zone into one dimension (the $x$ direction).
				(c) Valley-resolved Chern numbers emerge, due to the mass term and Landau levels (LLs, shown in grey).
				(d) Combined two valleys with the same $k_x$, quantized Chern number with valley-polarized edge states emerge.}
				}
\end{figure}

In conventional quantum valley Hall systems, the total Chern number vanishes, but the Berry curvature concentrates around the two inequivalent valleys with opposite signs, allowing one to define valley Chern numbers.
One can therefore assign a valley-resolved Chern number of $ \pm \tfrac12 $ to each valley, and the difference between them predicts helical edge states at a domain wall.
In the main text (Fig. 4), we can see that the interplay between the altermagnetic mass and the Dirac-type Landau levels leads to interesting Landau level asymmetry and valley-polarized edge states.
Here, we utilize the valley-resolved Chern numbers to reveal the underlying physics of this interesting phenomenon.

As discussed in the main text, the Brillouin zone of DFA contains four inequivalent valleys: two gapless cones with winding number $W = +1$ at $\Gamma$ and $M$, and two gapped cones with $W = -1$ at $X$ and $Y$ [Fig. \ref{FIGS7}(a)].
To analyze edge physics, we project the two-dimensional Brillouin zone onto one dimension (e.g., along $k_x$), which captures the topological spectrum of a nanoribbon [Fig. \ref{FIGS7}(b)].
Under this projection $\Gamma$ and $ Y$ map to $k_x = 0$, while $M$ and $X $ map to $k_x = \pi/a$.

In the gapped valleys $X$ and $Y$ the sign of the altermagnetic mass term $m_{\rm Am}(\cos k_x - \cos k_y)$ gives valley Chern numbers of \(\mathcal{C}_X= +\tfrac12 \) and \( \mathcal{C}_Y=-\tfrac12 \), respectively.
In the gapless valleys \( \Gamma \) and \( M \), the magnetic field leads to Landau levels, and the electron-like and hole-like branches contribute \( \pm\tfrac12 \) only depending on the field direction, as shown in Fig. \ref{FIGS7}(c).

Adding the two valleys that project onto the same \( k_x \) slice yields an effective one-dimensional valley Chern number: at \( k_x = 0 \) we obtain
\( C_{k_x=0} = C_{X} + C_{\Gamma} = -1 \) for the electron-like branch, and \( C_{k_x=0} =0 \) for the hole-like branch.
At \( k_x = \pi/a \) the situation is reversed: \( C_{k_x=\pi/a} = C_{X} + C_{M} = 0 \) for the electron-like branch and $C_{k_x=\pi/a}=+1$ for the hole-like branch, as shown in Fig. \ref{FIGS7}(d).
This momentum-dependent topology explains the asymmetric dispersion and fully valley-polarized quantum Hall edge states seen in Fig. 4.
At a deeper level, this reflects how the lattice regularization cancels the parity anomaly associated with individual Dirac cones.

According to a similar discussion, on the two sides of the domain wall around $x = 100a$ in Fig. 4(f), the left part and the right part has the valley-resolved Chern numbers $(\mathcal{C}_{\Gamma},\mathcal{C}_{X},\mathcal{C}_{Y},\mathcal{C}_{M})=\frac{1}{2}(-1,+1,-1,-1)$ and $\frac{1}{2}(-1,-1,+1,-1)$, respectively.
Along the domain wall, two parts have $\mathcal{C}_{k_y=\pi/a}=-1$ and $\mathcal{C}_{k_y=0}=-1$, resulting in the coexistence of valley-polarized edge states with the same Chern number.

\section{Emergence of net chirality in hexagonal lattice}

\begin{figure}
	\includegraphics[width=\columnwidth]{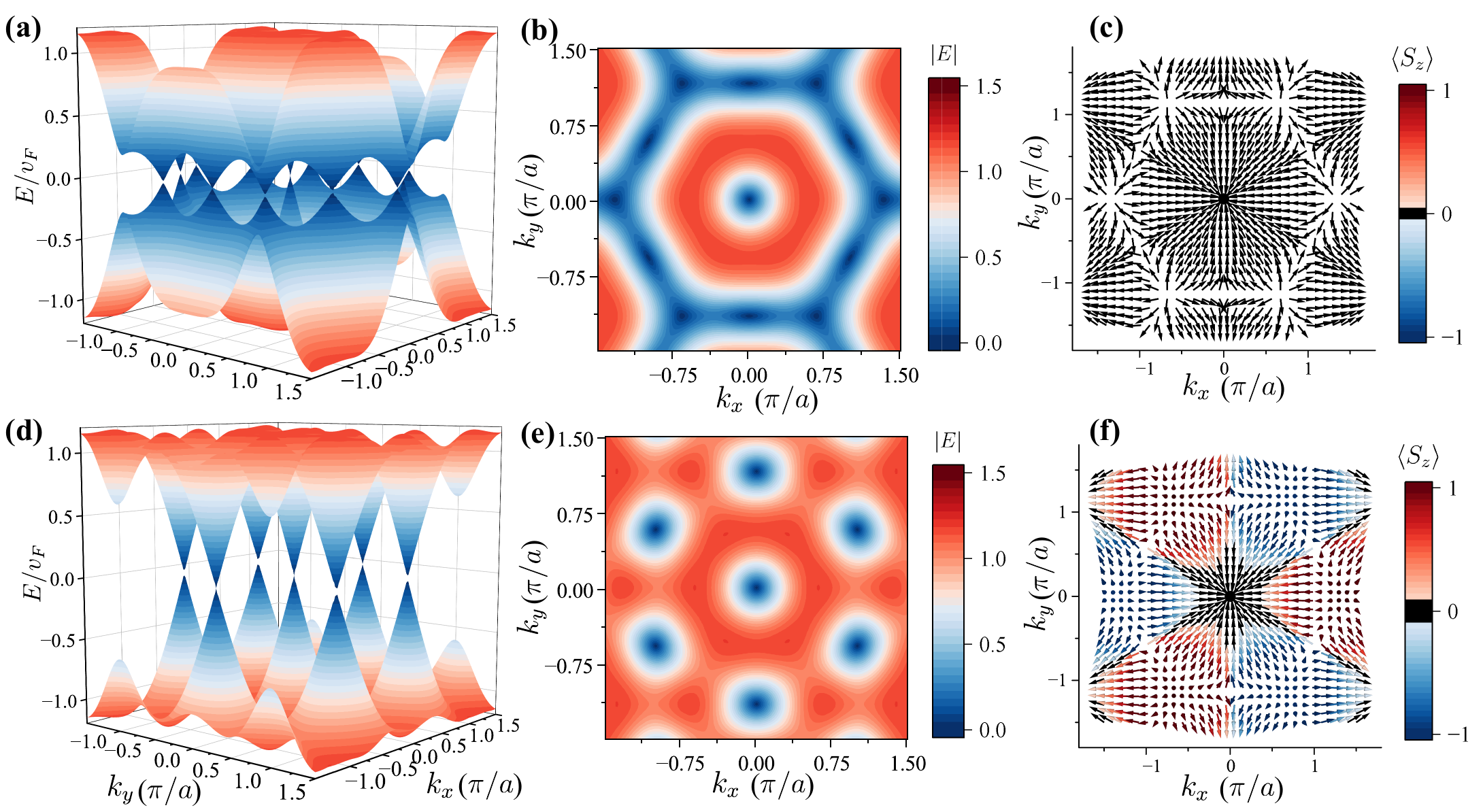}
			\caption{\label{FIGS8}
				\baselineskip 4pt {
				The case is in a hexagonal lattice.
				(a,b) and (d,e) are the band structures without and with $\mathcal{C}_6 \mathcal{T}$ altermagnetic mass $m_{\rm Am}^{h}$, respectively.
				(c) and (f) are the spin-textures without and with $m_{\rm Am}^{h}$, respectively.		
				In (a-c), six Dirac points emerge with canceled chirality.
				In (d-f), the altermagnetic mass $m_{\rm Am}^{h}=0.1v_F a_h$ gaps Dirac points at $K,K'$ and leads to net-chirality.}
				}
\end{figure}

In the main text, we found that the altermagnetic mass term with $\mathcal{C}_4\mathcal{T}$-symmetry can gap Dirac points with a single chirality in a square lattice, thereby giving rise to a net chirality.
In fact, the proposal we have put forward is not limited to the square lattice and $\mathcal{C}_4\mathcal{T}$ symmetry alone.

As an illustrative example, the hexagonal lattice provides an alternative platform.
In order to satisfy the symmetry of the hexagonal lattice, we perform central differences along three directions that form an angle of 120° with each other to discretize the Dirac Hamiltonian.
Therefore, $v_F(k_x \sigma_x + k_y \sigma_y)$ is replaced by:
\begin{equation}
H_{DF}^{h}(\bm{k})=\sum_{i=1,2,3} \frac{v_F}{a_h}\left[\sin (\bm{k}\cdot \bm{a}_i) (\frac{\bm{a}_i}{|\bm{a}_i|}\cdot \hat{x}) \sigma_x \right.+ \left. \sin (\bm{k}\cdot \bm{a}_i) (\frac{\bm{a}_i}{|\bm{a}_i|}\cdot \hat{y}) \sigma_y \right],
\end{equation}
where $\hat{x}$ and $\hat{y}$ are the unit vectors in the $x$ and $y$ directions, respectively.
$\bm{a}_1=(a_h,0)$, $\bm{a}_2=(-a_h,\sqrt{3}a_h)/2$, and $\bm{a}_3=(-a_h,-\sqrt{3}a_h)/2$ are the bond directions of the hexagonal lattice, with the bond length $a_h$.
As a result, six Dirac cones emerge [see Fig. \ref{FIGS8}(a,b)], as a result of the lattice regularization.
Three cones with winding number $W = +1$ appear at the $\Gamma$, $K$, and $K'$ points, and three with $W = -1$ at $M$ points [see Fig. \ref{FIGS8}(c)].

Next, we consider introducing an altermagnetic mass term based on $H^{h}_{DF}$.
Taking into account the symmetry of the hexagonal lattice, we use a mass term
\begin{equation}
H_{\rm DFA}^{h}(\bm{k})=H^{h}_{\rm DF}(\bm{k})+\frac{m_{\rm Am}^h}{a_h^2} \left[3-\sum_{i=1,2,3}\cos (\bm{k}\cdot \bm{a}_i)\right]\left[\sum_{i=1,2,3}\sin (\bm{k}\cdot \bm{a}_i) \right]\sigma_z,
\end{equation}
which is obtained by discretizing the Wilson mass with a $\mathcal{C}_6\mathcal{T}$ factor in a hexagonal lattice.
This altermagnetic mass term with $\mathcal{C}_6 \mathcal{T}$ symmetry gaps out the $K$ and $K'$ cones while leaving the others intact, yielding a net chirality and winding number of $-2$ [Fig. \ref{FIGS8}(d-f)].
Adding a trivial mass term $m\sigma_z$ to gap these net-chiral Dirac points will then lead to a quantized anomalous Hall phase, as in the square-lattice case.

Here, we further propose an experimentally feasible approach to realize the $\mathcal{C}_6 \mathcal{T}$-symmetric altermagnetic mass introduced in our model.
Recent theoretical and experimental studies have identified MnTe as a $g$-wave altermagnet, whose momentum-dependent Zeeman splitting in the $ k_z > 0$ sector respects the same symmetry \cite{MnTe1,MnTe2}.
This suggests a viable route: by placing a two-dimensional material in proximity to such an altermagnet, one can induce the desired $\mathcal{C}_6 \mathcal{T}$-symmetric mass term via interfacial exchange coupling or magnetic proximity effect.

\section{The prospects for future realization}

The central ingredient of our model, altermagnetism, has already been theoretically predicted in a wide class of materials and further shown evidences in several experiments, as documented in recent studies \cite{RuO21,RuO22,RuO23,dwave,MnTe1,MnTe2,LaMnO3,FeSb,CrSb}.
In fact, the first principle calculations and symmetry-based predictions have proposed countless altermagnetic candidates across various structural families.
For example, in 2025, a single article alone presents 50 new candidate materials \cite{new50}.
And the rules for searching for candidate materials based on \emph{ab initio} calculations and symmetry have already been established \cite{dwave}.
More specifically, among these candidates, there exist many $d$-wave and $g$-wave compounds that share key structural features with the lattice geometry used in our model \cite{RuO21,RuO22,RuO23,dwave,CrSb}.
Some of the materials, such as $\rm RuO_2$ \cite{RuO23}, $\rm MnTe$ \cite{MnTe1,MnTe2}, and $\rm CrSb$ \cite{CrSb}, have already shown signs of the expected altermagnetic spin splitting in the APRES experiments.

Furthermore, many of these materials incorporate heavy elements (e.g., $\rm Te$ \cite{MnTe1,MnTe2}, $\rm Ru$ \cite{RuO21,RuO22,RuO23}, $\rm La$ \cite{LaMnO3}, $\rm Sb$ \cite{FeSb,CrSb}), which may enhance the spin-orbit coupling.
This opens the possibility for realizing the band inversions and Dirac-like band structures with momentum-dependent spin-splitting, closely resembling the key ingredients of our theoretical model.
For example, the altermagnetic candidate material $\rm CoNb_3S_6$ has been regarded as a semimetal that hosts Dirac fermions \cite{RuO21,CoNbS}.
These considerations suggest that identifying material realizations with similar symmetry and magnetic features is a feasible and promising direction for future experimental efforts.

Compared with other schemes, either the Haldane model or the Wilson mass term currently shows no signs of being feasible for real system experiments.
The Floquet scheme requires external driving, which is not conducive to device applications.
The altermagnetic mass shows greater potential for device applications in topological electronics.
The momentum-dependent spin splitting resulting from its symmetry provides a broad prospect for the application of valley electronics.
